# GATOS IX: A Detailed Assessment and Treatment of Emission Line Contamination in JWST/MIRI Images of Nearby Seyfert Galaxies

Steph Campbell[1]*, David J. Rosario[1], Houda Haidar[1], Enrique López Rodríguez[16], Dan Delaney[11,12], Erin Hicks[9,11,12], Ismael García-Bernete[15], Miguel Pereira-Santaella[17], Almudena Alonso Herrero[13], Anelise Audibert[3,4], Enrica Bellocchi[7,8], Donaji Esparza-Arredondo[6], Santiago García-Burillo[2], Omaira González Martín[6], Sebastian F. Hönig[5], Nancy A. Levenson[14], Chris Packham[9,10], Cristina Ramos Almeida[3,4], Dimitra Rigopoulou[15,17], Lulu Zhang[9]

[1] *School of Mathematics, Statistics and Physics, Newcastle University, Newcastle upon Tyne, NE1 7RU, UK*
[2] *Observatorio Astronómico Nacional (OAN-IGN)-Observatorio de Madrid, Alfonso XII, 3, 28014 Madrid, Spain*
[3] *Instituto de Astrofísica de Canarias, Calle Vía Láctea, s/n, E-38205, La Laguna, Tenerife, Spain*
[4] *Departamento de Astrofísica, Universidad de La Laguna, E-38206, La Laguna, Tenerife, Spain*
[5] *School of Physics and Astronomy, University of Southampton, Southampton SO17 1BJ, United Kingdom*
[6] *Instituto de Radioastronomía y Astrofísica (IRyA-UNAM), 3-72 (Xangari), 8701 Morelia, México*
[7] *Departmento de Física de la Tierra y Astrofísica, Fac. de CC Físicas, Universidad Complutense de Madrid, E-28040 Madrid, Spain*
[8] *Instituto de Física de Partículas y del Cosmos IPARCOS, Fac. CC Físicas, Universidad Complutense de Madrid, E-28040 Madrid, Spain*
[9] *University of Texas at San Antonio, One UTSA Circle, San Antonio, TX 78249, USA*
[10] *National Astronomical Observatory of Japan, National Institutes of Natural Sciences (NINS), 2-21-1 Osawa, Mitaka, Tokyo 181-8588, Japan*
[11] *Department of Physics & Astronomy, University of Alaska Anchorage, Anchorage, AK 99508-4664, USA*
[12] *Department of Physics, University of Alaska, Fairbanks, Alaska 99775-5920, USA*
[13] *Centro de Astrobiología (CAB), CSIC-INTA, Camino Bajo del Castillo s/n, E-28692 Villanueva de la Cañada, Madrid, Spain*
[14] *Space Telescope Science Institute, 3700 San Martin Drive, Baltimore, MD, USA*
[15] *Department of Physics, University of Oxford, Keble Road, Oxford OX1 3RH, UK*
[16] *Department of Physics & Astronomy, University of South Carolina, Columbia, SC 29208, USA*
[17] *School of Sciences, European University Cyprus, Diogenes street, Engomi, 1516 Nicosia, Cyprus*
[17] *Instituto de Física Fundamental, CSIC, Calle Serrano 123, E-28006 Madrid, Spain*

**ABSTRACT**

Broadband mid-infrared (MIR) imaging with high spatial resolution is useful to study extended dust structures in the circumnuclear regions of nearby AGN. However, broadband imaging filters cannot distinguish dust continuum emission from emission lines, and so accounting for the emission line contamination becomes crucial in studying extended dust in these environments. This paper uses Cycle 1 MIR imaging from JWST/MIRI and spectroscopy from JWST/MRS for 11 local Seyfert galaxies, as part of the Galactic Activity, Torus and Outflow Survey (GATOS). Three of the objects (NGC 3081, NGC 5728, and NGC 7172) exist in both datasets, allowing direct measurement of the line emission using the spectroscopy for these objects. We find that extended MIR emission persists on scales of 100s of parsecs after the removal of contamination from emission lines. Further, the line contamination levels vary greatly between objects (from 5% to 30% in the F1000W filter), and across filters, so cannot be generalised across a sample and must be carefully treated for each object and band. We also test methods to estimate the line contamination when only MRS spectroscopy or MIRI imaging is available, using pre-JWST ancillary data. We find that these methods estimate the contamination within 10 percentage points. This paper serves as a useful guide for methods to quantify and mitigate for emission line contamination in MIRI broadband imaging.

**Key words:** galaxies: Seyfert – methods: observational – infrared: galaxies – galaxies: active – ISM: jets and outflows

## 1 INTRODUCTION

Our current understanding of massive galaxies places at the centre of each a supermassive black hole (SMBH). These black holes grow through the accretion of material from the central regions of their host galaxies, and through this process, they generate a vast amount of energy (Heckman & Best 2014; Morganti 2017). This energy is released both radiatively and kinetically, and injected into the material in the host galaxy – this is known as "feedback" and objects where this process is taking place are known as Active Galactic

* E-mail: steph.campbell@ncl.ac.uk

Nuclei (AGN). The process of black hole feeding and feedback is regarded as essential in our current understanding of galaxy evolution as a whole (Silk & Rees 1998; Di Matteo et al. 2005), with the energy from jets and winds launched by the AGN posited to regulate star formation within the galaxy on a global scale (see Veilleux et al. 2005; Fabian 2012; Veilleux et al. 2020; Harrison & Ramos Almeida 2024, for reviews). In order to understand the fundamental processes driving the feedback process, we must better understand the structure of the central few hundred parsecs.

From the observational side, constraints on the structures and processes at the centres of galaxies are hindered on two counts: first, the centres of galaxies are obscured by the gas and dust within the galaxies themselves; second, resolving structures on the scales of less than 100s of parsecs in external galaxies requires exceptional resolution (tenths of an arcsecond for galaxies ≤ 100 Mpc in distance). In the absence of direct imaging constraints on the geometry, studies of the spectral energy distributions (SEDs) of the unresolved nuclear emission (Ramos Almeida et al. 2009, 2011; Alonso-Herrero et al. 2011; Audibert et al. 2017; González-Martín et al. 2019a; García-Bernete et al. 2019; Esparza-Arredondo et al. 2019; Reyes-Amador et al. 2024) have found evidence for a central SMBH with an equatorial accretion disk, surrounded by an optically and geometrically thick dusty torus (Antonucci 1993; Urry & Padovani 1995), and this has been the prevailing torus geometry for the past few decades. This geometry explains the observed dichotomy between unobscured and obscured objects as being due to viewing inclination instead of any intrinsic physical differences, and so is known as the unified model of AGN. This model has been modified in recent years with the knowledge that the covering factor of the torus also plays an important role (see Ramos Almeida & Ricci 2017, for review), and varies across the population – with higher covering fractions (an intrinsic physical property) preferentially leading to obscured presentations (Ramos Almeida et al. 2011; Elitzur 2012; Audibert et al. 2017; Esparza-Arredondo et al. 2021).

Interferometry has allowed studies of the dusty torus structure in multiple regimes (see Hönig 2019). In the cold molecular regime ($T < 100$ K), the Atacama Large Millimeter Array (ALMA) has been instrumental in pushing the limits of observations, and has allowed the cold dusty torus to be fully resolved in a growing number of objects, putting constraints on the geometry, size, structure, and kinematics of the torus, and highlighting the presence of molecular outflows (García-Burillo et al. 2016; Alonso-Herrero et al. 2018; García-Burillo et al. 2019; Alonso-Herrero et al. 2019; Combes et al. 2019; García-Burillo et al. 2021; García-Bernete et al. 2021; Alonso Herrero et al. 2023; Esparza-Arredondo et al. 2025). In the hot regime ($T > 800$ K), VLTI/GRAVITY has probed the structure of hot dust around AGN and explored the sublimation region with sub-parsec resolution, finding little to no elongation in the hot structures (GRAVITY Collaboration et al. 2020b,a; Gravity Collaboration et al. 2024).

Moving back to cooler regimes in order to study the warm heated dust ($T \sim$ a few hundred K), infrared (IR) interferometry with the MID-infrared Interferometric instrument (MIDI) at the Very Large Telescope Interferometer (VLTI) has redefined our understanding of the geometry on these small scales. Modelling of this data suggests mid-infrared (MIR) emission in the polar direction in a number of objects in addition to the expected equatorial torus component. This is at scales and surface brightnesses which challenge the simplest unification model (assuming the same structure for all AGN) with an additional independent source of obscuration (Hönig et al. 2012; Tristram et al. 2014). Previous hints of extended MIR emission in the polar direction were attributed to dust heated in the narrow line region, or the illumination of dust at the inner edges of a conical void, and not associated with the torus (Bock et al. 1998; Cameron et al. 1993; Radomski et al. 2003). SED models which include this component have been successful in reproducing observations (Hönig & Kishimoto 2017; Alonso-Herrero et al. 2021; González-Martín et al. 2023), with most type-1 and luminous AGN being better described by disc+wind models, and less luminous type-2 AGN by clumpy torus models (González-Martín et al. 2019b; García-Bernete et al. 2022b). As it stands, the phase structure and content of AGN winds remains uncertain, but proving the presence of dust in these winds would be a helpful step towards understanding them better, and therefore understanding their dynamics (Soliman & Hopkins 2023; Sarangi et al. 2019) and their coupling to the ISM and effects on the host galaxy (see Ward et al. 2024, for recent simulation work exploring this).

Since this tentative glimpse of geometry beyond the classical torus, work has gone into developing physical models which can account for dust in the polar regions (Hönig & Kishimoto 2017; Hönig 2019). To aid in this, direct imaging of the polar structures would serve to provide definitive confirmation and offer constraints on these models. However, ground-based instruments with the power to resolve the $\approx 100$ pc scales expected (Williamson et al. 2020) have lacked the sensitivity required to image the faint extended structures, and suffer from the unstable PSF caused by the intervening atmosphere. Being space-based, the James Webb Space Telescope (JWST; Gardner et al. 2023) offers a stable PSF and also provides the sensitivity required to capture these low surface brightness structures. Warm dust of temperatures 150-300 K emits a blackbody continuum spectrum which peaks in the MIR, and is expected to be on the $\approx 100$ pc scale, which translates to an angular resolution on sub-arcsec scales. The Mid-Infrared Instrument (MIRI; Wright et al. 2023) on JWST offers imaging filters covering wavelengths from 5.6 to 25.5 microns, and so we can exploit JWST's sensitivity and stable PSF in order to open the window to direct images of the extended polar dust structures.

As part of the first cycle of JWST proposals, through the Galaxy Activity, Torus and Outflow Survey (GATOS; García-Burillo et al. 2021; Alonso-Herrero et al. 2021)[1], we capitalised on the power of JWST with a targeted imaging campaign to search for this extended MIR emission. The GATOS data are unique in using a special observing pattern in order to avoid saturation of the bright nuclei in nearby Seyfert galaxies. Initial results from this (Rosario et al. prep; Haidar et al. 2024) have revealed extended MIR emission in the polar direction around AGN; however, confirming that this is due to dust requires the treatment and mitigation of the issue of emission line contamination.

The imaging study uses broadband MIR filters, which capture the continuum emission expected from warm dust, and allow greater angular resolution (FWHM 0.328" in the F1000W filter) and sensitivity than spectroscopic instruments (close to 25% improvement, compare FWHM values in Figures 12 and 4 in Argyriou et al. 2023; Dicken et al. 2024, respectively), alongside a PSF which can be characterised and removed; however, these filters also cover some key emission lines originating from ionised and molecular gas (see McKinney et al. 2023, for a study of this effect in photometry of high-z dust obscured galaxies). Outwith the nucleus and regions affected by star formation, emission line regions in AGN are frequently connected to the outflows and found in the polar regions. As such, this can create false signals of polar dust if the polar MIR emission is incorrectly interpreted as being due to continuum emission, and create a stronger relationship between the dust distribution and the outflow regions than is truly the case. Further, the emission line re-gions have high surface brightness and are often patchy, which means the morphology seen in a broadband image can be greatly affected (D'Agostino et al. 2019; Harrison & Ramos Almeida 2024; Ward et al. 2024). As such, in order to conclude that the extended MIR structures observed are the result of dust as opposed to being generated by gas, we must account for the contribution from emission lines along the observed IR extended emission in the broadband images.

---

[1] https://gatos.myportfolio.com/

To this end, this paper combines the sample of nearby AGN imaged with JWST/MIRI as part of the GATOS imaging sample, with another sample within GATOS which has MIR integral field spectroscopy (IFS) data from the JWST/MIRI Medium-Resolution Spectrometer (MRS; Wells et al. 2015; Argyriou et al. 2023). These samples overlap, providing three targets for which there are both imaging and MRS data available. We refer to this subset of objects as the "shared sample". Thus the GATOS dataset offers a unique opportunity for this study (and many others), through the bespoke observing campaigns tailored to capturing regions near the bright central point source without saturation (García-Burillo et al. 2021; Alonso-Herrero et al. 2021; Rosario et al. prep). The MRS data allows for a direct measurement of the line contamination present in the broadband images over the overlapping fields of view, while the imaging allows testing and development of methods which can be used to mitigate and account for the contamination in objects where MRS data is not present (i.e. the rest of the imaging sample, future imaging campaigns). We produce synthetic line-free continuum images from the MRS data which act as contamination-free images tracing the continuum emission, and present the measurements of the spatially resolved line contamination across the galaxies. We then introduce, compare and evaluate methods to estimate and account for the contamination in the imaging data using ancillary data. This work is an important part of quality assessment of the GATOS imaging campaign, alongside quantifying this issue and introducing best practice mitigation for any future work which plans to use MIRI imaging.

In Section 2 we describe the JWST data used, alongside ancillary data we draw upon. Section 3 details the methods used to measure line contamination in objects with only spectroscopic data (Section 3.2), objects with both imaging and spectroscopic observations (Section 3.3) and introduces a method to estimate the contamination in objects with only JWST imaging data (Section 3.4). We present our results for one filter as a case study in Section 4.1, quantify the contamination across our sample in Section 4.2 and evaluate the accuracy of the estimation method in Section 4.3. We summarise our results in Section 5. This work made use of Astropy:[2] a community-developed core Python package and an ecosystem of tools and resources for astronomy (Astropy Collaboration et al. 2013, 2018, 2022), and reproject (Robitaille et al. 2024).

## 2 DATA

### 2.1 The Sample

Our sample combines objects selected for two existing GATOS projects. GATOS is using these samples of AGN with multiwavelength observations in order to better understand the processes and structures present in the central few parsecs of local AGN, and how these influence their host galaxies.

Eight nearby type 2 Seyfert galaxies were observed in five JWST/MIRI broadband imaging filters (F560W, F1000W, F1500W, F1800W, F2100W) as part of the JWST Cycle 1 GO proposal ID 2064

[2] http://www.astropy.org

(PI: D. Rosario). This work aimed to observe low surface brightness extended MIR emission in the inner few hundred parsecs of the AGN, which could be attributed to polar dust. The objects were selected as they showed evidence of polar dust structures on the scale of a few hundred parsecs in previous groundbased observations (García-Bernete et al. 2016; Asmus et al. 2016; Asmus 2019), were closer than 100 Mpc, and had a rich array of multiwavelength ancillary data. This subsample are described fully in Rosario et al. (prep).

Separately, six type 1.9/2 Seyfert galaxies were observed using the JWST/MIRI MRS instrument as part of the JWST Cycle 1 GO programme ID 1670 (PI: T. Shimizu and R. Davies). This work aimed to study PAHs and dust emission in warm outflows (García-Bernete et al. 2024b,a; Zhang et al. 2024b), alongside the ionised and molecular gas content (Hermosa Muñoz et al. 2024; Zhang et al. 2024a; Davies et al. 2024; Esparza-Arredondo et al. 2025). The objects were selected as a volume-limited sample of nearby (< 40 Mpc) luminous (log $L_{14-195keV}$ > 42.5) AGN from the *Swift*-BAT Ultra Hard X-ray Catalogue (Baumgartner et al. 2013). They were limited to those which have ionised outflow rates derived from optical spectroscopy, and so span a wide range of outflow rates within only a narrow range of luminosity. From here, the sample was finally restricted to AGN with host galaxies within a factor 2 of log $M_*$ = 10.43. This matching allows comparison of nuclear differences which are controlled for other factors. These data are described fully in García-Bernete et al. (2024b) and presented in Zhang et al. (2024a).

For the purposes of this paper, we do not intend for this to be a complete statistical sample of Seyfert galaxies, however we note that the sample still covers a range of X-ray luminosities (41.55 < log $L_X^{2-10keV}$ < 43.37 ; 42.58 < log $L_X^{14-195keV}$ < 43.59) and column densities (20.95 < log $n_H$ < 25.00), thus sampling a large region of the $n_H$-$L_X$ plane explored in García-Burillo et al. (2021) encompassing column densities representative of unabsorbed, absorbed, and Compton-thick systems; and luminosities congruent with the "blow-out" phase found in this plane.

Importantly, three of the objects (NGC 3081, NGC 5728, and NGC 7172, as seen in Figure 1) are shared between the imaging sample and the MRS sample, and so have JWST MIRI imaging in five bands, alongside JWST MRS spectroscopy. This allows in-depth analysis of emission line contamination in a spatially resolved manner. Through their inclusion in the imaging sample, these objects all have previous evidence of polar dust structures from groundbased observations. Throughout this work, these three objects are referred to as the "shared" sample.

Table 1 lists some key properties of the objects in the sample, and the data available. Below we summarise the observing strategies for the imaging and MRS data in the context of any details which are relevant to this work. For a full description, we direct the reader to the references above. We also describe any ancillary data which are used in this paper, including observations from the IFU Spectrograph for INtegral Field Observations in the Near Infrared (SINFONI; Eisenhauer et al. 2003; Bonnet et al. 2004) on the Very Large Telescope (VLT) and the Spitzer Space Telescope (Werner et al. 2004). Figure 1 shows RGB composite images from MIRI of the three shared objects in our sample, with the fields of view of additional datasets (MIRI/MRS, SINFONI and Spitzer) shown.

### 2.2 JWST MIRI Imaging

All eight galaxies targeted in the imaging sample were observed by the JWST/MIRI imager during JWST Cycle 1, with observations performed using the chosen 5 broadband filters (F560W, F1000W, F1500W, F1800W and F2100W). The observation process was tai-lored to the specific science case of searching for faint extended emission close to a bright nuclear point source which was also required to be unsaturated. As such, the dither sequence was selected with this in mind, alongside the choice to use a MIRI imaging sub-array for the observations to minimise readout time and mitigate saturation as much as possible. Further, changes were made to the standard JWST/MIRI data reduction pipeline in order to minimise saturated pixels further by including the first frame in the ramp fit which is otherwise usually excluded (see Haidar et al. 2024, for details of this).

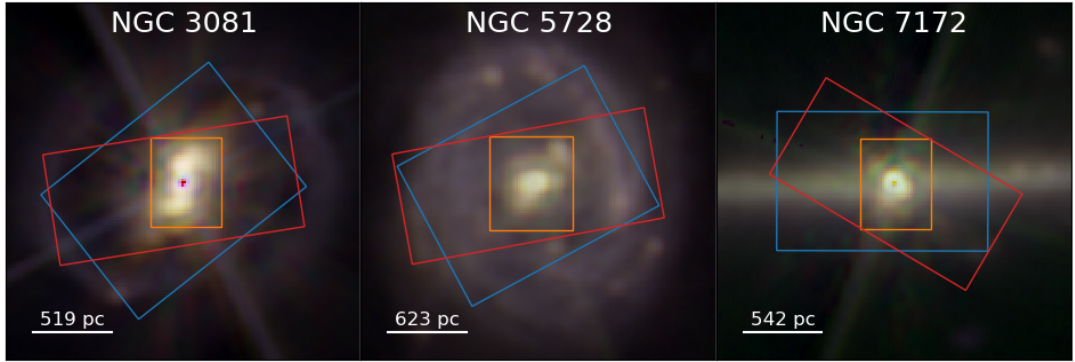

**Figure 1.** False colour JWST MIRI PSF-subtracted images of the three shared targets, NGC 3081, NGC 5728, and NGC 7172, covering a 15" × 15" FoV around the nuclear point source. In these RGB images, the blue, green and red channels have been assigned to the JWST MIRI F1000W, F1500W and F1800W filters respectively. Overlaid on the images are the fields of view covered by SINFONI (orange, 3.3" × 3.8"), JWST/MRS (blue, 7.3" × 9.0"), and Spitzer/IRS (red, 11.3" × 4.7").

| Object (a) | $D_L$ (b) | Redshift (c) | AGN Type (d) | $\log L_X^{14-195\mathrm{keV}}$ (e) | $\log L_X^{2-10\mathrm{keV}}$ (f) | $\log n_H$ (g) | Data Available (h) |
|---|---|---|---|---|---|---|---|
| ESO 137-G034 | 34.1 | 0.00914 | 2 | 42.59 | 42.80 | 24.30 | MRS Only |
| MCG-05-23-016 | 42.2 | 0.00849 | 1.9 | 43.57 | 43.288 | 22.18 | MRS Only |
| NGC 5506 | 26.7 | 0.00608 | 1i | 43.24 | 42.98 | 22.44 | MRS Only |
| NGC 3081 | 32.5 | 0.00798 | 2 | 43.18 | 43.02 | 23.91 | Imaging and MRS |
| NGC 5728 | 39.0 | 0.00932 | 2 | 43.27 | 43.21 | 24.13 | Imaging and MRS |
| NGC 7172 | 33.9 | 0.00868 | 2 | 43.18 | 43.37 | 22.91 | Imaging and MRS |
| NGC 3227 | 23.0 | 0.00376 | 1i | 42.77 | 42.84 | 20.95 | Imaging Only |
| ESO 428-G14 | 23.2 | 0.00566 | 2 | - | 41.55 | 25.00 | Imaging Only |
| NGC 2992 | 38.0 | 0.00771 | 1i | 42.58 | 42.67 | 21.72 | Imaging Only |
| NGC 5135 | 65.5 | 0.01369 | 2 | - | 43.26 | 24.40 | Imaging Only |
| NGC 4388 | 18.1 | 0.00842 | 2 | 42.61 | 43.04 | 23.52 | Imaging Only |

**Table 1.** Relevant properties for galaxies in the sample. The JWST/MIRI imaging observations were obtained through JWST Cycle 1 Program 2064, and those with JWST/MRS observations were part of JWST Cycle 1 Program 1670. (a): object names; (b): luminosity distance derived from redshift value; (c): redshift values from the NASA Extragalactic Database (NED; Helou et al. 1991); (d): AGN types from Burtscher et al. (2016), 1.9: as defined in Osterbrock (1981), 1i: broad hydrogen recombination line(s) detected at infrared wavelengths, 2: no broad hydrogen recombination line detected; (e): log 14-195 keV X-ray luminosity in ergs/s, derived from *swift*/BAT flux corrected for absorption; (f): log 2-10 keV X-ray luminosity in ergs/s, derived from flux corrected for absorption (flux for MCG-05-23-016 sourced from Mattson & Weaver 2004); (g): X-ray column density in $cm^{-2}$; (h): JWST data available for use in this work.

Additional deviation from the standard pipeline is employed in finding an astrometric solution to the field. The default method to register the astrometry of each set of images before combining them into science-ready products uses point sources (stars or distant galaxies) within the frame. As our images capture the central regions of nearby galaxies, background distant galaxies are not detectable, and stars are not visible in the MIR range. As such, the default method is not suitable for these targets. Instead, the astrometry was tweaked manually to align the central point source to within sub-pixel scales.

Matching this centroid position to the centroid position found in ALMA data allows absolute astrometric solutions to be achieved.

### 2.3 JWST MIRI/MRS MIR spectroscopy

All galaxies in the MRS sample used here were observed as part of JWST Cycle 1. The MRS IFU covers a spectral range from 4.9 to 28.1 µm, with data across 4 channels: Ch1 (4.9-7.65 µm), Ch2 (7.51-11.71 µm), Ch3 (11.55-18.02 µm), Ch4 (17.71-28.1µm). Within each of these channels, light is split to three different grating settings (short, medium and long; as seen in Figure 3), and as such one MRS observation is received as 12 separate cubes after reduction. The data were reduced primarily through the standard MRS pipeline procedure (e.g. Labiano et al. 2016, and references therein). This used pipeline release 1.11.4 and the calibration context 1130. An extra step was added (as described in Appendix A of García-Bernete et al. 2024b) to handle hot/cold pixels in the data before the final cubes were produced. The full process is described in detail in García-Bernete et al. (2022b, 2024b) and Pereira-Santaella et al. (2022).

In this work, we assume that there are negligible astrometric mis-alignments across all 12 cubes. Earlier MIRI spectroscopic studies that have employed these cubes have reported small variations in the centroid of the continuum emission with wavelength (e.g. Section 4.2 of Davies et al. 2024), but these can be interpreted as changes in the small-scale dust structure around the AGN rather than astrometric jitter.

### 2.4 VLT/SINFONI NIR spectroscopy

Archival K-band data from the IFU Spectrograph for INtegral Field Observations in the Near Infrared (SINFONI; Eisenhauer et al. 2003; Bonnet et al. 2004) on the Very Large Telescope (VLT) were compiled for a sample of 89 nearby Seyfert galaxies as part of the LUNIS-AGN catalogue (Delaney et al. 2025). These observations employed adaptive optics for improved resolution. This sample covers all the galaxies used in this work. The data were reduced by the SINFONI Data Reduction Software (Abuter et al. 2006), before the IDL routine LINEFIT (Davies et al. 2011) was used to fit single gaussians to the lines of interest in each spaxel. This gives the data products used directly in this paper, which are the [Si vi], Brγ and $H_2$ linemaps. For this work, which focuses on the region of the ionisation cones, an extra data processing step was performed to clean the noise in the outskirts of the maps. Any spaxels in the linemap with a signal-to-noise ratio less than 1.5 were masked, and the linemaps were smoothed using a gaussian kernel equivalent to the FWHM of the JWST filter in which they are going to be used.

### 2.5 Spitzer/IRS spectroscopy

In this work, we rely on existing MIR spectroscopy from the *Spitzer* InfraRed Spectrograph (IRS; Houck et al. 2004) for measurements of the emission lines that are the primary contaminants in the broadband MIR images. Spectra from the Spitzer/IRS high-resolution arms ($R \sim 600$) are needed for accurate measurements of the line fluxes, though they only cover the wavelengths of 9.9–37.2 $\mu m$, smaller than the range of wavelengths spanned by the MIRI filters of the GATOS imaging programme. In Section 3.4, we present an approach that can be used to estimate the fluxes of key lines not covered by the Spitzer/IRS high-resolution spectra. Fully-reduced archival high-resolution spectra are available for all targets in the imaging sample from the Cornell Atlas of Spitzer/IRS Sources (CASSIS; Lebouteiller et al. 2011)[3]. For our purposes, we use CASSIS spectra from the "full extractions", which gives us the total lines fluxes captured by the Spitzer/IRS high-resolution slit apertures. We elect to use this option in order to offer a maximal estimate for the contamination, as the full extraction includes flux from a wider field of view (FoV) which more closely matches that of the SINFONI data (see Figure 1), whereas the alternate "optimal" extraction is matched to PSF size. Because the apertures are different for short and long wavelengths, and also due to the substantial increase of the PSF width of Spitzer over the wavelength range of the spectra, we note that the spatial regions sampled by the spectra are not the same across wavelength. Lines at longer wavelengths (18.7-37.2 $\mu m$) are extracted from a larger part of the target galaxies (an aperture of $11'' \times 22''$) compared to lines at shorter wavelengths (9.9-19.6 $\mu m$) which are extracted from a smaller aperture (an aperture of $5'' \times 11''$). All of our lines fall within the range of the smaller aperture.

---

[3] http://cassis.astro.cornell.edu/atlas

## 3 METHODS

In this paper we set out to quantify the line contamination in broadband MIRI images of the circumnuclear regions of our sample. We define line contamination in a pixel of a particular MIRI broadband image as the percentage of the total throughput-weighted flux in that pixel which arises from line emission. Beyond characterising the line contamination, we also aim to "decontaminate" the images, in order to produce a "line-free" image. This is an image which shows how the object would appear in a given filter if it were observed without the emission lines present, and is referred to as a "decontaminated image". While this removes the emission lines, we note that the broad silicate (e.g. at 9.7 microns) and PAH features are not removed, however this is not important as silicates and PAH are both still associated with dust. These decontaminated images are used to draw qualitative conclusions as to whether the extended MIR emission is due to line-free dust emission. In Lopez-Rodriguez et al. (prep), these are used to explore the physical properties of this dust.

To aid the reader, our three methods are shown schematically in Figure 2, and described in the following Sections. We begin in Section 3.1 by explaining how we generate synthetic broadband images from the MRS data, decompose these into line-only and line-free emissions in Section 3.2, mitigate the effects of a central point source in Section 3.2.1, effectively creating a decontaminated image directly. This allows contamination to be quantified through the comparison of the line emission to the full synthetic image. In order to improve upon the results of this through better resolution and mitigated PSF effects, we introduce a method in Section 3.3 to use a combination of imaging and MRS data for the 3 objects which are shared across the datasets. The resulting output from this method is referred to as the "reference" decontaminated image/contamination map. The reference results are then used to evaluate the accuracy of an estimation method, described in Section 3.4, which is used to decontaminate and estimate the contamination in objects for which only MIRI imaging data is available. In all cases, the end products are a map of contamination, as defined above, and a "decontaminated" line-free map. This is achieved through the subtraction of the contamination from the original image (whether a MIRI image or a synthetic broadband image from MRS).

The results of these methods are presented and evaluated in Section 4.

### 3.1 Generating Synthetic Broadband Images from MRS Spectra

The default MRS reduction pipeline produces a separate cube for each of the four channels and each of the three gratings, a total of 12 cubes. When attempting to create synthetic images that sample a range of wavelengths under a broadband filter function, one will run into a situation where different wavelengths are covered by different cubes. In the top panel of Figure 3, different colours are used to show spectra from the short (green), medium (orange), and long (pink) grating settings of Channel 2, each of which comes from a different cube. The filter function of the MIRI F1000W imaging filter (grey dashed line) has some transmission across wavelengths from all of these cubes.

In general, different cubes may not sample the same FoV, or even be centered on the same aim point. Therefore, we developed a procedure to generate an aligned and registered cube for all wavelengths spanned by a given MIRI broadband filter. Such filter-specific cubes were used to perform synthetic photometry and produce synthetic broadband images of our targets from their MRS spectral datafiles.

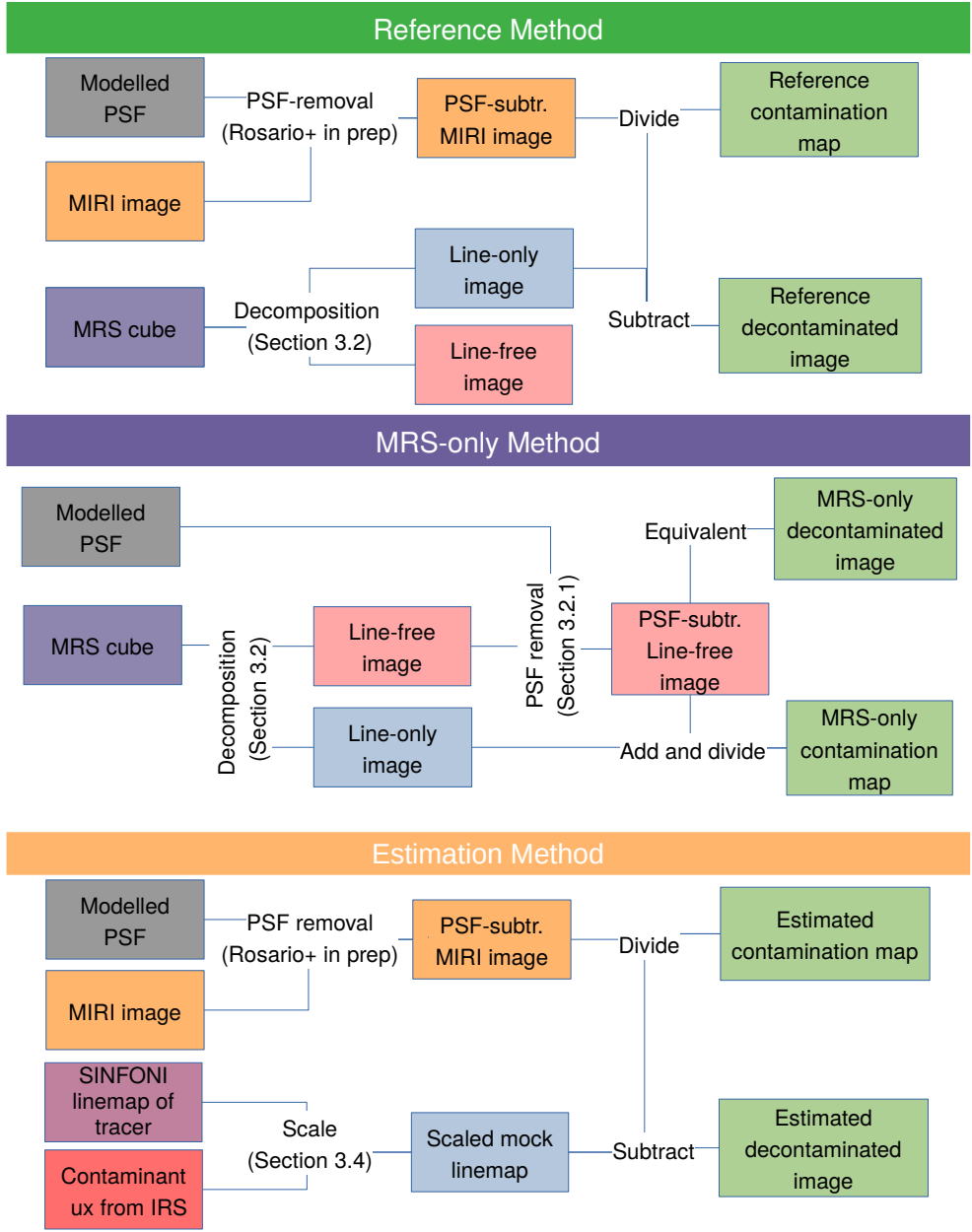

**Figure 2.** Schematic outlining the three different methods employed to generate contamination maps and decontaminated images depending on data available. Green boxes highlight the outputs which are shown in Figure 4 for the three shared objects.

We used the Astropy-affiliated Python package `reproject` to generate the filter-specific cubes for a given MIRI broadband filter. `reproject` uses world coordinate system (WCS) information, stored in the FITS file headers of the MRS pipeline outputs, to map one image to the sky coverage of another at a subpixel level, conserving the surface brightness of the mapped image. The code handles differences in the geometry of the pixels in the two images from shifting, rotation, and distortion, as long as this information is encapsulated by the WCS keywords of the associated FITS files.

We first defined the span of the filter as the complete range of wavelengths over which the filter transmission function is greater than 1/1000 of its peak value. The filter functions we used were tabulated by the University of Arizona MIRI team before the launch of JWST[4], but these are consistent with the Cycle 1 performance of MIRI before the recent degradation of the long-wavelength sensitivity of the detector.

Using header information from the FITS files for all MRS cubes, we identified the specific cubes that overlapped in wavelength with the span of the filter. For all the filters from the GATOS MIRI imaging programme, this process picked between two and three cubes, continuous in wavelength, which we will call "spanned cubes" henceforth.

From the geometrical information stored in the header of each cube, we identified the cube with the smallest FoV, and used its WCS information to define the reference frame for the final filter-specific output cube. Looping over each wavelength increment of the spanned cubes, we used `reproject` to map the slice at that wavelength, essentially a very narrow-band image, on to the reference WCS frame. We employed the most accurate spherical polygon intersection algorithm in `reproject`, which conserves surface brightness. As the pixel sizes of the final cube could be different from that of the constituent cubes, we scaled the mapped slice by the ratio of the spaxel areas from the original and reference cubes to ensure that the final cube conserved flux, rather than surface brightness.

[4] http://svo2.cab.inta-csic.es/svo/theory/fps3/

There is a small degree of wavelength overlap between cubes that are adjacent along the spectral direction. We use these overlap regions to estimate the small flux scalings that are needed to bring all spectra to the same flux scale of the reference cube, individually in each spaxel. This generates a continuous spectrum in each spaxel that covers the full span of the chosen filter. After applying these flux scalings, we stacked the resultant mapped slices into the final filter-specific cube.

To create the final synthetic broadband image, we started by converting the cubes from native units of MJy/sr into working units of erg s$^{-1}$ cm$^{-2}$ $\mu$m$^{-1}$ pixel$^{-1}$. After interpolating the tabulated filter functions to the wavelength array of the final cube, we numerically integrated, over wavelength, the product of the filter function and the spectrum in each spaxel of the cube. We divided these filter function-weighted fluxes by the equivalent integral of just the filter function itself.

The final product of this procedure is an image of a target that synthesises an observation taken by the MIRI imager in a particular filter, derived from, and covering the FoV, of the MIRI/MRS cubes of the target. Note that the pixel scale of this synthetic image is not the same as the MIRI images, but corresponds to that of the spanned MRS cube with the smallest FoV.

## 3.2 Measuring Line Contamination in Synthetic Broadband Images Using MRS Data

In order to quantify line contamination in a broadband image synthesised from MRS data, the MRS spectrum from the "spanned cube" must be decomposed into continuum emission and line emission. We demonstrate our approach in Figure 3, where the upper panel displays the integrated MRS spectrum for NGC 5728, between ~ 7.5 and ~ 12 microns in different colours for each MRS channel. The throughput for the F1000W MIRI imaging filter is shown by the dashed line. This illustrates the issue of line contamination in the F1000W ($10\mu m$) band. Considering a central circular aperture of 2 arcseconds, we find a significant amount (varying across objects but reaching up to 25%) of the flux is coming not from the continuum but from line emission from [S IV], [Ar III], and H$_2$ 0-0 S(3). This is further exaggerated in the F1000W band due to silicate absorption dampening the continuum contribution from the nucleus. To account for the contamination due to the emission lines, the data can be decomposed into a line-only spectrum and a line-free spectrum. There are also PAH emission features found at 8.6$\mu m$ and around 11.2$\mu m$ which contribute to the flux in the broad band (seen in Figure 3). These are left as part of the line-free spectrum due to their minimal contribution within the filter (the feature at 8.6$\mu m$ overlaps with only a very low transmission part of the filter, while the complex at 11.2$\mu m$ emits only low levels of flux), and the additional complexity in modelling them (see Figure 3 of García-Bernete et al. 2022a, for example).

To perform this decomposition, we first define narrow continuum windows around each of the contaminating emission lines, as shown by the faint grey regions in the upper panel of Figure 3, between 1000 and 2000 km s$^{-1}$ from the line centre. We use these windows to derive straight line approximations to the continuum underlying the emission lines. We create the line-free spectrum by replacing the small regions covered by the emission lines (1000 km s$^{-1}$ around the line centres, highlighted by the darker grey regions in Figure 3) with these continuum approximations. Following this, we create the pure line emission spectrum by subtracting this line-free spectrum from the integrated spectrum. The lower panel of Figure 3 shows the integrated line-free spectrum in blue and the integrated line emission spectrum in green for illustrative purposes. In practice, this procedure is applied to the spectrum from each MRS spaxel of the spanned cube in order to decompose a line-only and line-free cube covering the wavelengths encompassed by the broadband filter. After this, we follow the methodology outlined in Section 3.1 to create equivalent bandpass-weighted line-only and line-free images from the line-only and line-free cubes, respectively.

### 3.2.1 PSF Treatment in MRS Data

A major complication in an accurate determination of line contamination from MRS data is the influence of the bright nuclear point source, which, when combined with the complex JWST PSF, diffracts light into the rest of the MRS FoV. As the nuclear point source is strongly continuum-dominated, the diffracted light, if not removed, artificially lowers the estimate of line contamination in regions close to the nucleus. (The results of our processing without addressing the PSF in MRS data are presented in Appendix B for completeness and comparison).

The modelling and mitigation of the PSF in bright point sources from MRS cubes is a challenging topic and the field is only just beginning to develop methods for this. González-Martín et al. (2025) introduced a successful method to decompose an MRS cube into a point source cube and an extended cube. That work focused on recovering the point source emission in a spectrally resolved manner for use in torus modelling, as opposed to recovery of the broadband extended emission which we desire here. While successful for its desired goals, the method leaves strong residuals which make the interpretation of small-scale morphology in the extended emission difficult. Given the priorities of our work compared to González-Martín et al. (2025), we pursue a different technique described below.

As we are estimating contamination from synthetic images in a broadband MIRI imaging filter, we employ the well-tested MIRI imager models of the JWST PSF from the software package WebbPSF (Perrin et al. 2012). We build on the methods that have been developed by our team for MIRI imager studies (Rosario et al. prep). From JWST commissioning studies, it is known that the MRS PSF has a width that is about 25% larger than the diffraction-limited PSF of JWST (Argyriou et al. 2023). We recreate this effect in WebbPSF by including a larger Gaussian jitter in the PSF models, tuned to produce a final PSF that has the same width as the MRS PSF at the central wavelength of the respective imaging filter.

To subtract the nuclear point source from the line-free synthetic images, we adopt an approach similar to that used for the MIRI images, outlined in Section 2.2. This provides PSF-subtracted line-free synthetic images. Visual inspection shows that these images are free of most of the extended diffraction spikes, but there are cases of over- or under-subtraction in the PSF core. This is entirely expected - the core is particularly susceptible to the accuracy of the PSF model. As we are interested in the line contamination in the extended emission, we proceed with our current PSF subtraction approach, while cautioning the reader that residual artifacts will be visible in the MRS-only contamination maps.

The various synthetic images generated for all three shared objects in the F1000W filter are shown in Figure 4. Those generated for the other filters are presented in Appendix C.

### 3.2.2 Creating MRS-Only Contamination Maps

To generate a map of line contamination in each spaxel from the MRS data, we add the line-only synthetic image to the PSF-subtracted line-free synthetic image, in order to produce an approximation of the original broadband synthetic image without the central point source. We use this along with the line-only synthetic image to produce the MRS-only contamination map, following the definition our definition from Section 3. These are presented for all three shared objects in the F1000W filter in the right-most column of Figure 4. Those generated for the other filters are presented in Appendix C.

While removing the PSF of the point source, our method still leaves strong residuals which are visible in the maps, and hinder understanding of the circumnuclear structures which are targeted. The spatial resolution of MRS observations is approximately 25% worse than that of the MIRI images, which further reduces the suitability of this method for studying the small extended structures. Therefore, whilst theoretically MRS data allows "true" measurement of the contamination, the issues with PSF and resolution make them sub-optimal for these purposes. As such, we pursued an improved method in Section 3.3.

### 3.3 Measuring Line Contamination for Objects with MRS Spectroscopic Data and MIRI Imaging Data

In order to achieve a measurement of the line contamination which is free from the challenges of mitigating bright point source PSF effects in MRS cubes, we use a combination of MRS spectroscopic data and MIRI imaging data to create our reference contamination maps. For this, we use a set of nuclear point source-subtracted broadband images available for these targets along with the MRS-derived line-only maps. A detailed description of the point-source subtraction procedure is given in Rosario et al. (prep), and key details are summarised below. For an assessment of the performance of this method, an applied example is found in Haidar et al. (2024).

As expected for AGN, the objects exhibit extremely bright point sources, which lead to strong effects from the complex PSF of the MIRI imager. The PSF features a bright core alongside a hexagonal ring structure, with a set of 6 diffraction spikes from the hexagonal mirror, and a further 4 diffraction spikes in a crosshair pattern from the detector which is more pronounced at shorter wavelengths. All of these vary according to wavelength and detector position. A method to model and subtract this was developed and employed to mitigate some of the effects.

The PSF-modelling tool WebbPSF (Perrin et al. 2012) was used to create accurate models of the PSFs for the images in each of the five MIRI bands, including the effects of dithering and the time-dependent wavefront error of MIRI. While we note the differences between the WebbPSF model and the PSF generated by the MRS instrument in Section 3.2.1, it is documented in Dicken et al. (2024) that the MIRI PSF is diffraction limited, and the WebbPSF model is tested and evaluated as an accurate model. One dimensional azimuthally-averaged surface brightness profiles of the images, centred on the AGN, were fit using a combination of the PSF model and an underlying Sersic model for the extended MIR emission, within a region corresponding to 3 times the 80% encircled energy radius of the respective PSF. The resultant scaling of the nuclear point source from the fits is used to subtract the PSF model from the nuclear position of the image, leaving only the extended emission. These fits are primarily governed by the Airy structure of the bright nuclear point source. However, residual diffraction structure can remain in the PSF-subtracted images because the extended emission in the MIR is also often centrally-concentrated. While this method is not perfect as per the limitations detailed in the paper above, and seen through the artifacts remaining in the images in Figure 1, this does allow study of the underlying extended emission which was otherwise hidden by the PSF.

| Band | Range [$\mu$m] | Line | $\lambda$ [$\mu$m] | IP [eV] |
|---|---|---|---|---|
| SINFONI | 1-2.5 | [Si VI] | 1.96 | 166.7 |
| | | H$_2$ 1-0 S(1) | 2.12 | - |
| | | Br$\gamma$ | 2.17 | 13.3 |
| MIRI F1000W | 9.0-10.9 | [Ar III] | 8.99 | 27.6 |
| | | [Fe VII] | 9.53 | 99.1 |
| | | H$_2$ 0-0 S(3) | 9.66 | - |
| | | [S IV] | 10.51 | 34.8 |
| MIRI F1500W | 13.5-16.6 | [Ne V] | 14.32 | 97.1 |
| | | [Ne III] | 15.56 | 41.0 |
| MIRI F1800W | 16.5-19.5 | H$_2$ 0-0 S(1) | 17.04 | - |
| | | [S III] | 18.71 | 23.3 |
| MIRI F2100W | 18.5-23.2 | [S III] | 18.71 | 23.3 |

**Table 2.** List of emission lines detected in the wavelength ranges covered by different instruments and filters, alongside their ionisation potentials (IPs). H$_2$ information from Black & van Dishoeck (1987), other species information compiled from Gelbord et al. (2009), Sturm et al. (2002) and May et al. (2018).

Another option to recover the underlying nuclear structure is deconvolution – Leist et al. (2024) explores this method using NGC 5728 from the GATOS imaging sample (see this study for a comparison of the methods). For the study of the extended structure, subtraction is adequate, and so all MIRI images in this paper which are described as "PSF-subtracted" have been achieved using the method from Rosario et al. (prep).

The central point source is continuum-dominated, with very little line emission associated with the nuclear spectrum. This can be verified from the lack of PSF structure in the line-only maps produced (see Figure 6 and those included in Appendix C). Assuming that the point source is a pure continuum emitter, the nuclear point-source subtracted from the MIRI images only removes the central continuum. Consequently, subtracting the synthetic line-only image from the nuclear point source-subtracted MIRI image should give a line-free image, in which the effects of the bright central PSF are strongly mitigated as the PSF-subtraction for MIRI imaging results in minimal residual artifacts, thus yielding a more accurate map of the extended continuum emission. This method also allows the superior resolution and sensitivity of MIRI imaging to be preserved.

Given the benefits of combining higher resolution point source-subtracted MIRI imaging and MRS line maps, we treat this as the "reference" decontaminated image. Such reference images can be produced for the 3 objects which are shared across the MIRI imaging and MRS datasets. Using these "reference" results as a benchmark to test against, we develop a method for estimating contamination using only imaging data. If we are able to accurately estimate the contamination from emission lines in MIRI images without the need for expensive MRS spectroscopy, these images can be used as a reliable tool to explore the structure of the extended dust in the circumnuclear regions of our AGN.

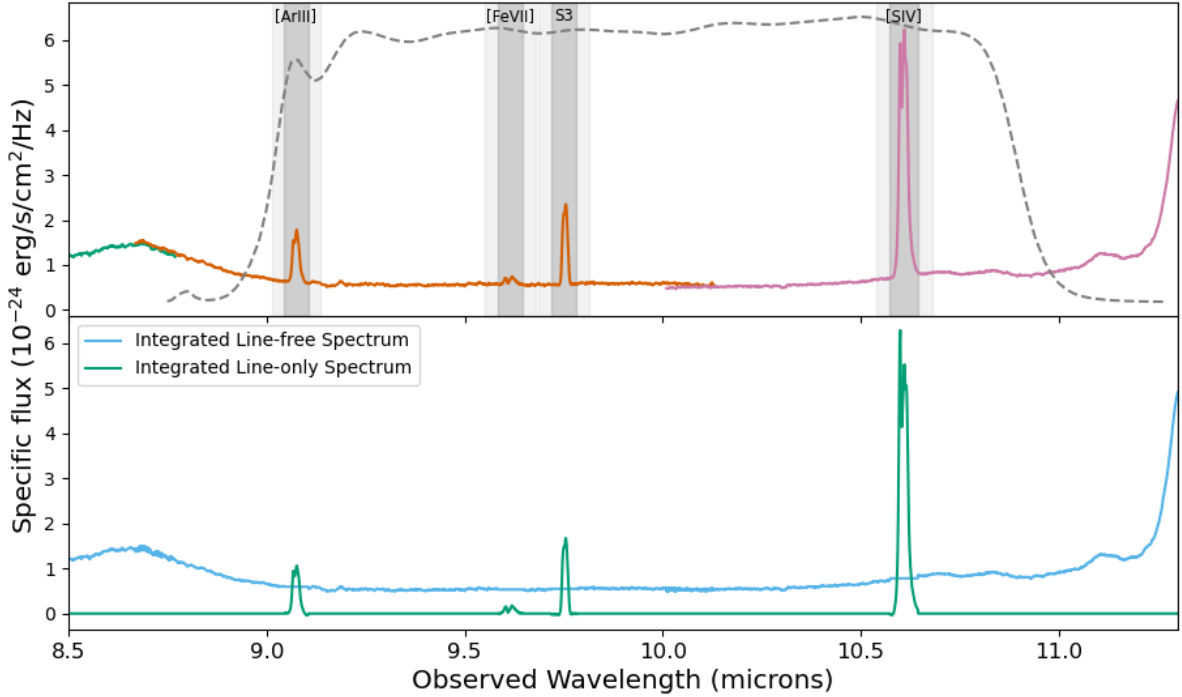

**Figure 3.** Top: spatially integrated spectrum from the MIRI MRS "spanned cube" for NGC 5728, showing the different gratings in green, orange and pink; overlaid in dashed grey is the MIRI F1000W broadband filter throughpass curve. Emission lines which contribute to flux within the filter range are marked and labelled. Bottom: the same spatially integrated MIRI MRS spectrum, this time decomposed into the emission line-only spectrum (green) and the line-free spectrum (blue) to illustrate the method described in Section 3.2.

### 3.4 Methods for Estimating Line Contamination from Imaging Data Using Ancillary Data

In Haidar et al. (2024), a method was introduced which allows an estimated measurement of the line contamination in images for objects which do not have MRS data. This method requires a spatial tracer of an emission line which is contaminating the chosen filter (i.e. a map of an emission line which has similar ionisation potential), and a flux estimation of this emission line. For example, in the F1000W broadband filter, we know the primary contamination is due to [S IV]. Without MRS data, we do not have a spatial map of where within the FoV this emission is coming from; however, a species with similar ionisation potential would be expected to be found tracing similar regions. From SINFONI IFU observations (described in Section 2.4), we can produce a map of the spatial distribution of Br$\gamma$, and from Spitzer/IRS an unresolved spectrum is available, from which we measure the total flux from [S IV] within a similar FoV (see Figure 1). These measurements are achieved by subtracting a linear continuum and fitting a single gaussian profile to the lines. Combining these, we are able to create a mock [S IV] map by assuming the [S IV] will trace the same morphology as the Br$\gamma$ and scaling this to match the total [S IV] flux within the Spitzer/IRS FoV.

From this principle, we can create mock line maps for other species which are found to be contaminating the filter. Continuing with F1000W as an example, Figure 3 shows $H_2$ 0-0 S(3) produces emission in the filter at 9.66$\mu m$. SINFONI provides a map of $H_2$ using the $H_2$ 1-0 S(1) line, which we assume traces the same spatial distribution as $H_2$ 0-0 S(3). The flux of S(3) is not available from Spitzer/IRS data, though $H_2$ 0-0 S(2) is present. Further the MRS data contains both $H_2$ 0-0 S(2) and $H_2$ 0-0 S(3). As such, the MRS sample were used to measure the ratio between the S(3) and S(2) fluxes in a 3" FoV, again by fitting a single gaussian to the line profile. This gave an average value of $R_{\frac{S(3)}{S(2)}} = 3.41 \pm 1.31$. Assuming this to be representative across the sample, therefore the flux of $H_2$ 0-0 S(2) measured in the Spitzer/IRS data is scaled by $R_{\frac{S(3)}{S(2)}}$ in order to give an estimated flux we would expect from $H_2$ 0-0 S(3). This value is then used to scale the flux in the SINFONI $H_2$ 1-0 S(1) map in order to produce an estimated mock $H_2$ 0-0 S(3) map which is used to estimate the contamination.

Finally for F1000W, there is an [Ar III] line. This line is typically subdominant to the [S IV] line and $H_2$ 0-0 S(3) line, and so is less important to model. As the *Spitzer* high resolution data does not cover the wavelength of this line, we must again use the MRS sample in order to derive a ratio in order to estimate this. As such, we measure the ratio of [Ar III] to [S IV] in the full MRS sample and find that $R_{\frac{[Ar III]}{[S IV]}} = 0.45 \pm 0.23$. This is assumed to be representative across the sample, and so the estimated mock map is created by scaling the Br$\gamma$ map to have a total flux equal to the [S IV] flux multiplied by $R_{\frac{[Ar III]}{[S IV]}}$. This is of course a simplistic assumption – ratios such as these are affected by density of medium, temperature, and other factors, which will inherently vary spatially across the FoV. For the purpose of this estimation in the absence of MRS data, we proceed with these caveats and added errors in mind.

The flux in F1000W also includes a contribution from an [Fe VII] emission line. This is a weak line; we consider its effects to be negligible and leave it out of our estimates.

Thus a full estimated linemap for the F1000W band is generated by the sum of:

- the [S IV] mock – traced by Brγ from SINFONI and scaled to the [S IV] flux measured in Spitzer/IRS
- the H2 0-0 S(3) mock – traced by H$_2$ 1-0 S(1) from SINFONI and scaled to the H$_2$ 0-0 S(3) flux which is in turn recovered through the conversion factor from MRS data converting the H$_2$ 0-0 S(2) flux measured from Spitzer/IRS
- the [Ar III] mock – traced by Brγ and scaled to the flux of the [Ar III] line, which is recovered from the ratio between [S IV] flux and [Ar III] flux in the MRS sample

Note that there are corrections needed to estimate the contributions of two of these three lines, which are based on empirical scalings from the MRS data. These scalings have a sizable variation across objects averaged over, which limits the accuracy of this proposed estimation method. However, as one may gather from Figure 3, the majority of the contamination in F1000W arises from the [S IV] line which is directly measurable from Spitzer/IRS high-resolution spectra.

The full procedures used for estimates of lines in other filters are described in Appendix C, but the general principles followed are: SINFONI maps of either Brγ, [Si VI], or H$_2$ are used as spatial tracers and assigned based on the IP of the contaminating line, the flux is scaled to the flux measured in Spitzer/IRS if the contaminating line is covered by this instrument, if not then a ratio to a related line is measured from MRS spectra for the MRS objects and the average of this is taken as a conversion factor.

In Appendix A we provide Figure A1, which shows the morphology of the three tracers used from SINFONI and compares them to the morphology from the MRS data of the contaminating lines in the F1000W filter. Our method includes many assumptions which should be kept in mind as caveats. As already mentioned, assuming constant conversion factor across the FoV is simplistic given the many varied conditions which are covered. There is also the issue of dust attenuation which could add additional discrepancies in the spatial distributions recovered for the contaminants from the tracer species. For example, in using [S IV] to trace Brγ, regions of higher attenuation will result in lower fluxes at lower wavelengths, and thus Brγ will be less bright in the more attenuated regions. Using [S IV] as a tracer for Brγ means the scaled maps will have more flux present in these regions than in reality. Considering the morphology of attenuation maps in these types of objects (see Prieto et al. 2014, for a map for ESO428-G14 among other objects), higher attenuation is seen in the central regions. For our final results, a measurement of emission line contamination, higher attenuation in the centre will lead to an overestimate of contamination (which, for this science case, is preferential to an underestimate, as discussed further in Section 4.3). Additionally, Haidar et al. (2024) considers the issue of attenuation and extinction, and considers it to vary an insignificant amount in their object to impact results. Other objects seen in Prieto et al. (2014) exhibit similar magnitudes of variation, and thus the same conclusion can be assumed for our sample. Indeed, we see in Figure A1 that, despite all these caveats and simplistic assumptions, the morphological similarities between the maps of the tracers vs the contaminants provide peace of mind that the overall distributions of the emission lines are captured. We also note the background surface brightness levels in SINFONI which are notably higher than that in the MRS data – this leads to high levels of estimated contamination in areas with low flux, which we discuss further.

## 4 RESULTS AND DISCUSSION

In order to showcase our methods fully, we elect to use the F1000W filter as a case study, and step through the methods used on the three objects with both MRS and imaging data available (the "shared" objects; namely NGC 3081, NGC 5728, and NGC 7172). This filter is chosen as it is an effective "worst case scenario" for line emission contamination. This is due to the strong emission lines present within the band alongside the effects of silicate absorption, as well as the capture of lines which are not represented in the IRS data, and so require additional steps to estimate. Additionally, this is the filter of interest for the science case of identifying extended dust emission. Section 4.1 showcases the results of our methods in this filter. Section 4.2 gives some statistics and general characterisation of the contamination across the sample. And finally Section 4.3 evaluates the accuracy of the methods. For the reader's reference, we include the full results for all galaxies in all filters in Appendix C, using whichever method is suitable based on the data available.

### 4.1 Line Contamination in the F1000W Filter: A Case Study

As seen in Figure 3, the F1000W filter is contaminated by multiple lines: [S IV], H$_2$ (0-0) S(3), and [Ar III]. Figure 4 shows the results of using all 3 methods described in Sections 3.2, 3.3, and 3.4 on our 3 objects with MIRI imaging and MRS spectroscopy – NGC 5728, NGC 7172, and NGC 3081 – in order to quantify and mitigate this contamination. For each object, the top row showcases the results of using the "MRS + imaging" method – highlighting the use of the PSF-subtracted MIRI image and the line-only map from MRS in order to create a "reference" decontaminated image and contamination map. This is the most robust method for decontamination, and in all cases we see residual extended structure in the MIR continuum-emission.

This method is compared with the lower two rows which demonstrate the results achieved if we use only one of the two datasets. The middle row illustrates the results from using only MRS data, where the F1000W broadband image is synthesised as per Section 3.1, decomposed into a line-only image and a line-free image (as described in Section 3.2), and the bright central point source is removed as per 3.2.1. Here the line-free image serves as the "decontaminated" image. We see strong residual artifacts caused by oversubtraction of the central point source PSF structure. These regions of oversubtraction show artificially weak line-free emission, and so result in erroneous peaks in the contamination maps. In Appendix B, we show the results if the MRS PSF was not removed. Without the removal of the PSF from the bright point source, the extremely strong PSF in the MRS data dilutes the extended structure in which we are interested, and artificially lowers contamination estimates in the central regions. Even with the artifacts resulting from oversubtraction, we can see the underlying extended MIR continuum structure is recovered from the MRS data with a similar morphology to the reference method in the top row. The superior PSF-subtraction results, alongside the improved spatial resolution, highlights the strength of the reference method compared to the MRS-only method.

The bottom row demonstrates that the quantification of line contamination using the estimation method which combines JWST imaging with ancillary data, as described in Section 3.4, can come close to recovering the measured contamination found using the reference method. Comparing the contamination maps in the bottom row with those of the top row, it is clear that the estimation method overestimates the level of contamination. However, despite this overestimation, we see extended structure on scales of 100s of parsecs in the decontaminated map which is consistent with the structure seen using our reference method.

We evaluate and discuss the accuracy of the estimation method in more detail in Section 4.3, while the physical properties of these extensions are explored in Lopez-Rodriguez et al. (prep).

NGC 5728 is by far the most contaminated object in the sample of 3 shared galaxies, in part due to silicate absorption dampening the continuum, with emission line contamination reaching in excess of 50% of the flux in the far parts of the extended MIR region. We also note the peaks of the contamination running along the leading (clockwise) edges of the x-shaped cone structure, this is in line with results seen in Davies et al. (2024) which posit that the ionised emission seen on these edges are due to shocks as the rotating material is intersected by the ionisation cone.

NGC 3081 displays the most striking extended MIR emission of the three objects, with an S-shaped structure oriented north and south, aligned with the known ionisation cone. This shape is echoed in the emission line maps, however when the amount of contamination is measured it is found to be consistently around 15% or lower in the regions of extended emission, regardless of the method used. As such, when the contamination is removed, this extended emission structure remains clearly visible. This object has a strong PSF in the MRS data (see Figure B1 in Appendix B), which would obscure the the detail of the structure in the line-free map, however our PSF-subtraction method allows us to recover the underlying structure (albeit, with strong artifacts present). By using the MRS data in tandem with the imaging data in our reference method, we are able to recover this structure more clearly. The estimation method also does well in recovering the qualitative structures with fewer PSF-subtraction artifacts. In regions of low flux in the MIRI image, the higher noise in the background regions of the SINFONI data causes anomalously high and inaccurate contamination signals. We note that these are high contamination levels on low amounts of flux, and thus do not affect the conclusions about the extended MIR structures.

NGC 7172 shows hints of extended MIR structure to the southwest of the nucleus, though this is uncertain as it aligns with the crosshair pattern which remains as an artifact in the MIRI images and MRS lacks resolution to confirm it. This feature echoes the direction and scale of extended cold dust emission found in Alonso Herrero et al. (2023) using ALMA, and is in the region impacted by the outflow. NGC 7172 also has a very bright point source which is seen in the strong nuclear PSF in the unmitigated MRS data (see Figure B1 in Appendix B). Using the unmitigated MRS data leaves results which are fully lost in the PSF structure. With the PSF removed as per Section 3.2.1, we see some contamination recovered to the southwest of the nucleus, however this is severely affected by an artifact. As seen by the reference contamination map, there is contamination present here. This contamination is aligned with the potential extension (as this is aligned with the ionisation cone) but remains around 10% of the flux in both the estimated and reference maps, and so the emission lines cannot fully account for the extended emission which persists in the decontaminated images. We note again that in regions of low MIRI flux, the noisier SINFONI map causes regions of apparent high contamination, though these are on the disk of the galaxy and not related to the extended MIR regions which we are attempting to study.

Overall, despite contamination being present in the extended structures, when it is correctly measured and accounted for, we see in the decontaminated images (in all methods) that the extended MIR emission is still present – this is in line with the results seen in the Haidar et al. (2024) study of ESO 428-G14 which used the estimation method. This is true for all three galaxies shown here. In NGC 5728, the contamination is significant and reaches above 50%. As seen in Figure 4, this removes the MIR blob to the northwest, and reduces the extension seen to the southeast, however there is still extension present here and to the west. The contamination in NGC 7172 and NGC 3081 is much lower than that found in NGC 5728, reaching only 20% and 30% respectively. As seen already, this is at its strongest towards the outskirts of the extended emission regions, and the removal of the contamination still leaves extended MIR emission in the polar direction of both (though on very small scales in NGC 7172).

Qualitatively, the results from the three methods are not dissimilar. However, the significant and complex nuclear PSF which was present in the MRS data has resulted in strong artifacts from oversubtraction in our maps. This is discussed further in the following section where we evaluate and compare the methods.

### 4.2 General Characterisation of Contamination in the Sample

In Figure 5, we present the contamination maps and decontaminated images using the reference method for all three shared objects across the four filters. As the percentage contamination maps show, emission line contamination is only a serious issue at shorter wavelengths, with the maximum contamination in a pixel decreasing from 50% in F1000W, to 30% in F1500W, 15% in F1800W, and 10% in F2100W. While from a contamination point of view, F2100W would seem superior for the study of warm dust emission, the considerably worse resolution and point spread function mean that small scale extended features are lost. Equally, F1000W gives the best resolution to detect these small scale MIR structures, but contamination is strongest in this filter and so hardest to mitigate for. We do note that the structures are not completely lost in the longer wavelength filters, and the persistence of the extended structures in these much less contaminated wavebands further backs up their existence. Even using the less robust methods, we still see this trend of lower contamination in longer bands.

In Table 3 we present the percentages of flux-weighted average contamination within the ionisation cone in an annulus from 0.25-2 arcsec, for each of our objects, using all three methods. We see here that NGC 5728 is the most contaminated in all filters, using all methods. Crucially, the large variation in these values from object to object shows that emission line contamination cannot be generalised across objects, and requires explicit treatment in every case.

### 4.3 Characterising the Accuracy of the Methods

Shifting focus now to the accuracy of the methods using only one of the JWST datasets, Figure 6 displays the emission line map generated through the estimation method compared to the emission line-only map extracted from the MRS data for our three shared objects. The final column displays the percentage residual of the difference between the two, divided by the total flux in each spaxel – ie the difference in the percentage of contamination in a pixel from the estimated linemap compared to the MRS linemap. In all three cases, outer regions display overestimates in the estimated emission line map (redder pixels), as the SINFONI data – even after cleaning and masking – contains more noise in these regions. In the areas of interest where there is potential extended MIR emission, the percentage difference is close to zero near the nucleus aside from in NGC 5728. Looking again to Table 3, we can consider the flux-weighted average contamination levels within 2 arcseconds within the ionisation cone, and this time compare different methods for each object. In the 10μm band, we see

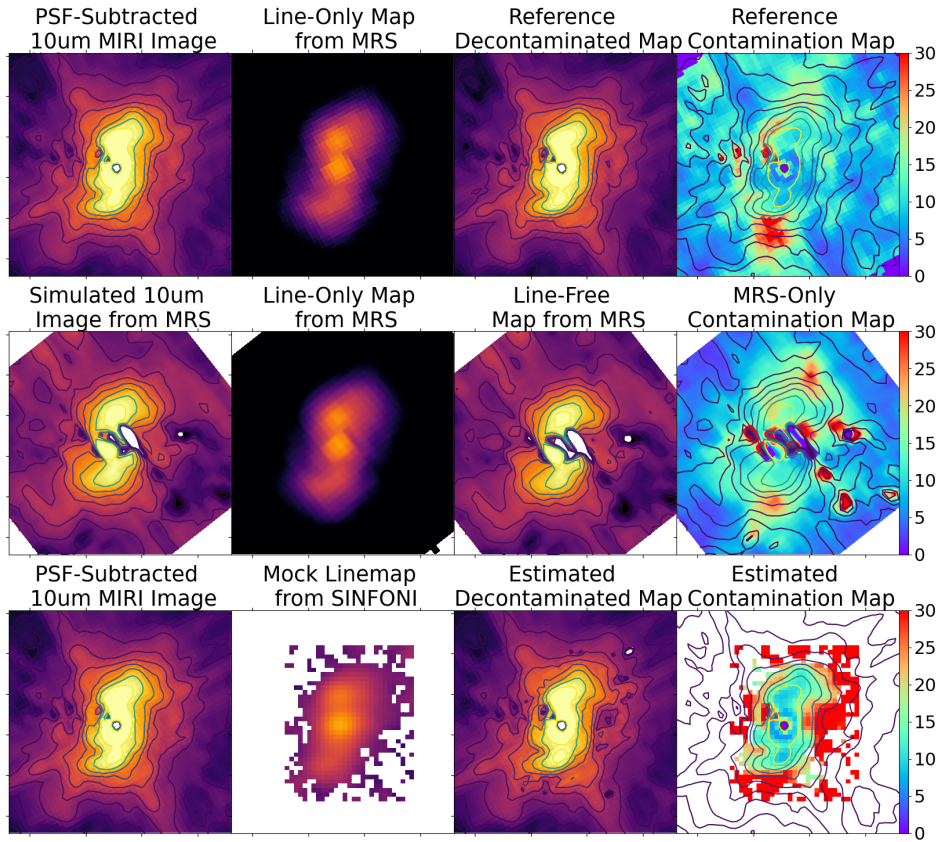

((a)) NGC 3081

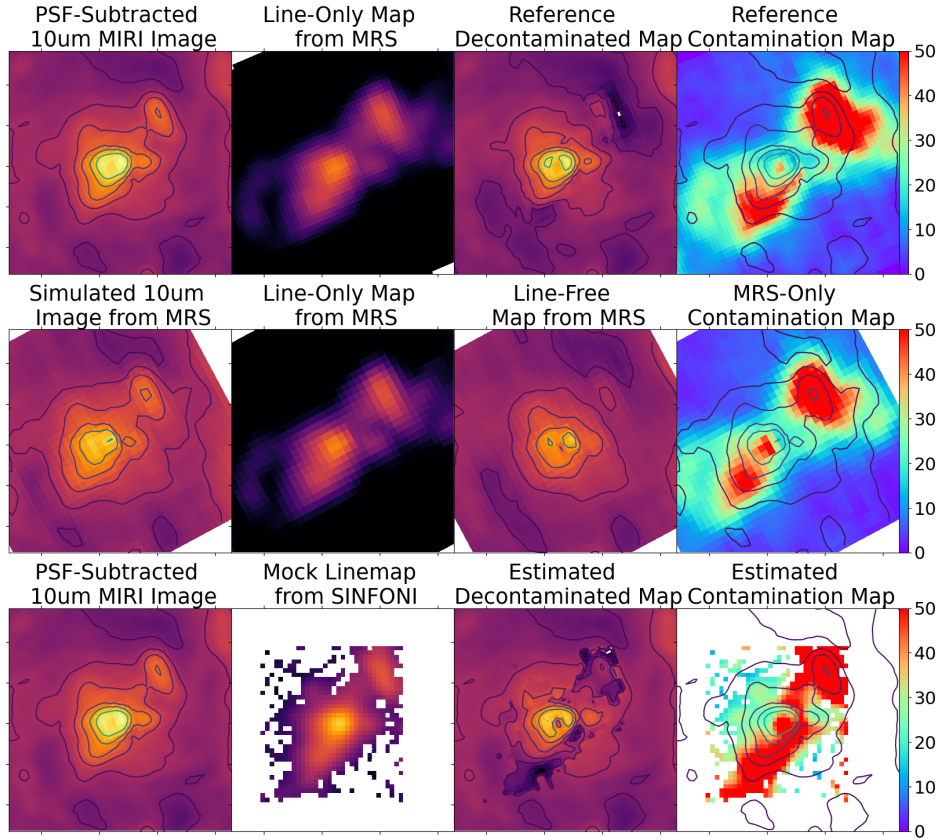

((b)) NGC 5728

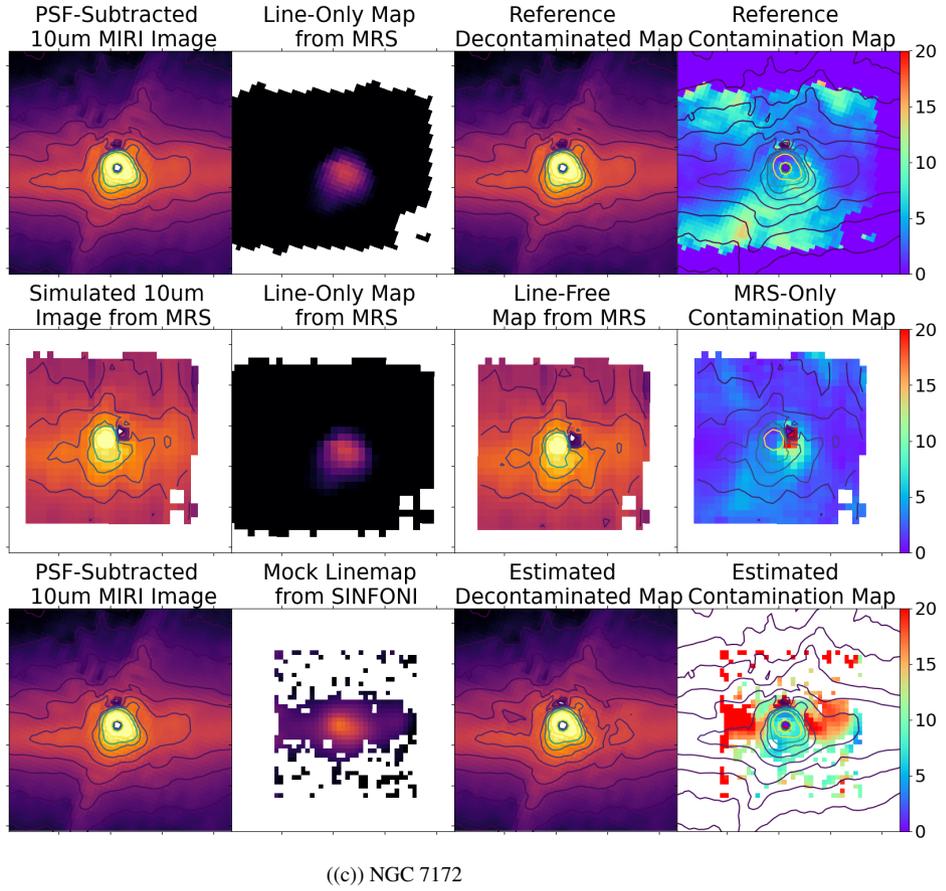

((c)) NGC 7172

**Figure 4.** A summary of all three methods used on the three shared objects in F1000W – NGC 3081, NGC 5728, NGC 7172. The top row shows the reference method: the PSF-subtracted MIRI image is combined with the line-only MRS map, subtracting this gives the decontaminated map, and the contamination percentage map is the amount of flux from line emission compared with that in the PSF-subtracted image. The middle row shows the MRS-only method: the full image is synthesised from the MRS data, and can be decomposed into line-only emission and line-free emission. In this case, the synthetic line-free map is effectively the decontaminated image. The contamination map is the percentage of flux in the line-free map compared with the synthetic image. The bottom row shows the estimation method: The PSF-subtracted MIRI image is combined with a mock linemap generated from SINFONI data, subtracting this linemap from the image gives the estimated decontaminated map, and the contamination map is created through the ratio of the linemap flux to the image flux. The contours in these images, in all cases, represent the flux. The contours in the first column shows the flux in the image. The contours in the third column are that of the flux in the decontaminated map. The contours in the fourth column are those of the flux in the initial image.

| Object | Method | F1000W | F1500W | F1800W | F2100W |
|---|---|---|---|---|---|
| NGC 3081 | Reference | 11.58 | 5.13 | 1.29 | 0.78 |
|  | MRS | 18.65 | 5.78 | 1.46 | 0.93 |
|  | Estimate | 14.46 | 3.75 | 0.68 | 0.44 |
| NGC 5728 | Reference | 34.91 | 12.87 | 5.11 | 3.42 |
|  | MRS | 38.97 | 19.73 | 6.03 | 3.08 |
|  | Estimate | 44.70 | 14.40 | 4.21 | 2.89 |
| NGC 7172 | Reference | 5.03 | 4.03 | 1.95 | 0.99 |
|  | MRS | 3.75 | 17.72 | 3.85 | 1.27 |
|  | Estimate | 11.20 | 3.07 | 1.87 | 1.00 |

**Table 3.** Average percentage of flux-weighted contamination within the ionisation cone in an annulus from 0.25-2 arcsec (see Figure 7) across objects and methods.

in all objects that the estimation method overestimates the contamination compared to the reference method, while the MRS method can both over and under estimate the contamination. This is analysed further in Figure 7, where the flux-weighted average contamination in the F1000W band is measured for the three methods (reference method in orange, MRS method in blue, and estimation method in green) in annular sectors to recover the radial profile of contamination within the known ionisation cone of the objects. Also shown in this Figure is the surface brightness profile in these same regions (grey dashed line), which highlights how the estimated contamination increases sharply in regions of low MIR surface brightness (i.e. beyond the regions of interest for extended nuclear MIR emission) due to noisy SINFONI maps. We fade away regions beyond where the average surface brightness reaches below 10% of the peak value. Figure 7 shows more clearly that the estimation method (green line) either matches (in the case of NGC 3081) or overestimates the con-tamination, while the the MRS method (blue line) overestimates the contamination in the centre, trending closer to matching the reference method at greater radii (except in NGC 7172, where the MRS method underestimates at all radii). The overestimation of the estimation method is only minor (to the same degree as the MRS underestimation in NGC 7172) however for the purposes of determining whether an MIR structure is dusty or if it is due to emission line contamination, an overestimate is preferred. An underestimate of the contamination could lead to a false positive signature whereas an overestimate of the contamination has less impact on the interpretation of remaining MIR emission in a decontaminated image – a few percent overestimate is unlikely to erase an extension fully whereas a few percentage points underestimated would leave detectable extensions which may be falsely attributed to dust. While the MRS method does not consistently result in underestimation, the unpredictability of whether an underestimate or overestimate has occurred means the conclusions from this method would be inconsistent.

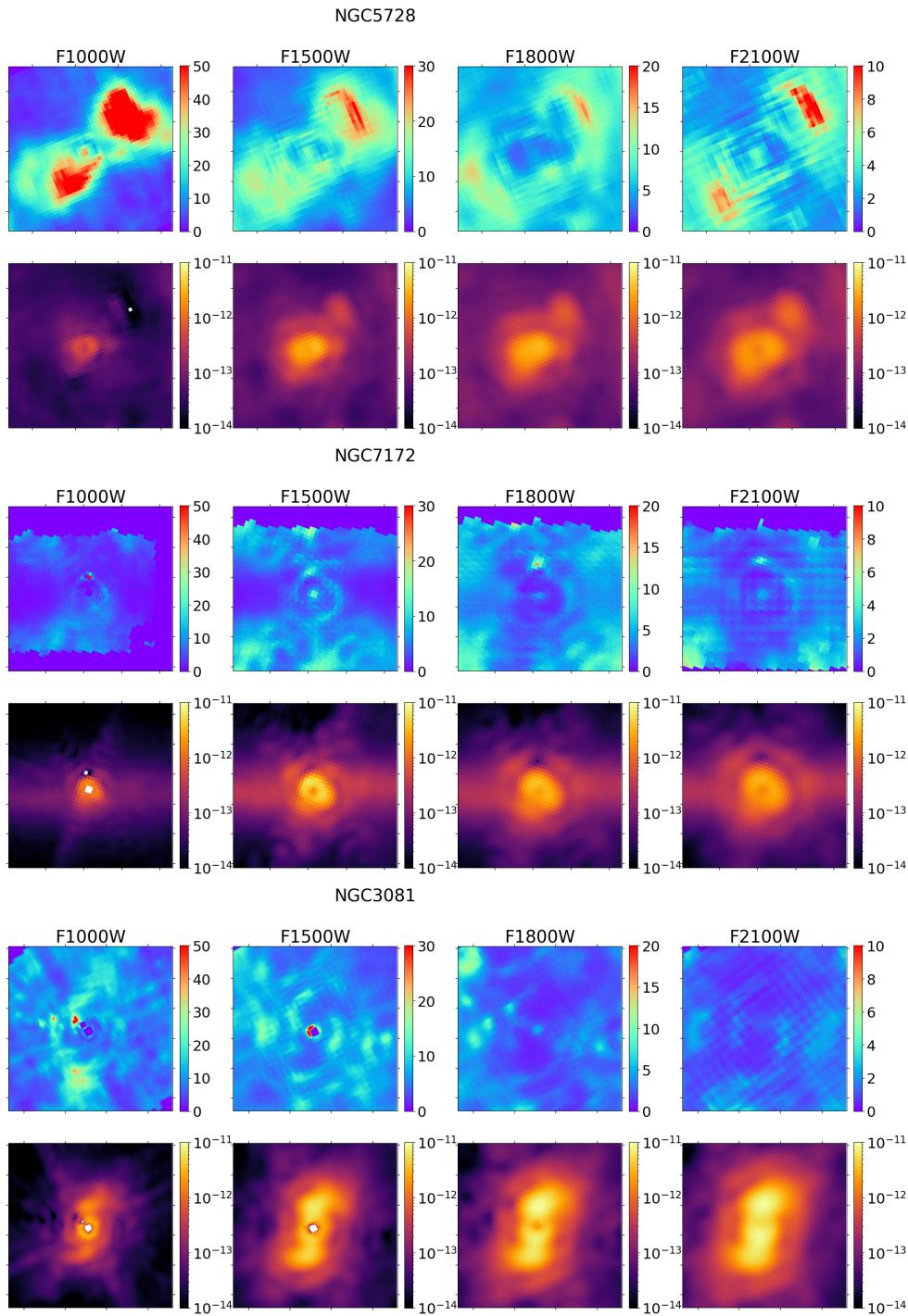

**Figure 5.** Contamination maps (top row) and decontaminated images (bottom row) generated using the reference method for the three shared objects shown across all filters.

Given the analysis above, this work advocates that the reference method which combines imaging with MRS data is the best option to quantify the contamination in JWST broadband data and recover the line-free image. However, despite offering improved accuracy over the other methods, the resulting decontaminated images and contamination derived from this method are still affected by multiple limitations. As we are combining datasets from the MIRI imager and the MRS, it is essential that the fluxes are accurately calibrated and the backgrounds are subtracted in both instances. As the FoV of both the image and the MRS data cover the nuclear region and do not extend beyond the host galaxy, this means background subtraction is not straightforward. Haidar et al. (2024) details the background subtraction method used for the imaging data, and Rosario et al. (prep) describes how this is applied across the imaging sample. For the MRS sample, dedicated background observations taken far from the galaxy are used for the background subtraction. Another limitation arises from the combination of data with different PSFs. While we have already discussed the PSF effect from the strong nuclear point source, the PSF of both instruments causes different levels of "spreading out" of compact structures. As the MRS has a wider PSF than the MIRI imager, any light from compact structures which is included in the MRS synthetic line-only image will be more spread out than the corresponding structure in the image. As such, when the MRS synthetic image is divided by the MIRI image in order to estimate the contamination in the reference method, the contamination in compact structures will be underestimated due to flux being more concentrated in the MIRI image, while this same light is more spread out in the synthetic line emission image from the MRS. All this said, the reference method still offers the best treatment of the contamination as the MRS data allows direct measurement of the emission line flux, while the imaging allows better resolution and PSF mitigation.

Nevertheless, we note that it is observationally expensive to obtain both MIRI imaging and MRS data for objects such as these, and so this paper aims to offer an exploration and evaluation of the other options. As seen in Section 4, using MRS data alone does allow direct measurement of the contamination, however it suffers from artifacts from the complexity of PSF mitigation for this instrument, reducing the levels of trust in morphology retrieved for the nuclear structures. If these objects were not so inherently bright at the nucleus, this method would offer a good measurement. Alternatively, the use of imaging alone cannot disentangle the light coming from line emission, but our estimation method shows that combining imaging with ancillary data such as SINFONI to create mock linemaps can recover the underlying line-free emission in a way which avoids the issues of the MRS PSF and comes closer to qualitatively reproduce what is found from in the reference results. The method does overestimate the contamination by a few percent, but for the science purposes outlined here we favour an overestimation over an underestimation. As such we conclude that for objects where ancillary SINFONI data is available, the use of imaging is preferential to the use of MRS data (where it is not possible to get both, in which case this is superior).

We note that the estimation method introduced here may also be suitable for similar mitigation of emission line contamination in other instruments such as NIRCam, and for other science cases which require study of underlying continuum emission in broadband filters. We offer this method as a base to be built upon in future, including exploring the applicability to other instruments and wavelength regimes. Further, we also note that the improvements in modelling and mitigating the MRS PSF introduced in González-Martín et al. (2025) may yet improve the results acquired from using the MRS-only method and we look forward to seeing this in the near future.

## 5 CONCLUSIONS

In this paper we have explored the issue of emission line contamination in the use of broadband JWST/MIRI imaging for the study of extended continuum emission in local Seyfert galaxies from the GATOS sample. Using different methods, we have quantified the contamination in a spatially resolved manner, and generated decontaminated images of the objects. This has allowed us to discern whether extended MIR emission is from dust, or line emission. We have made use of a set of three objects with both high-quality MIRI/MRS spectroscopic data and multi-band MIRI imaging data. This has allowed us to directly measure the flux due to emission lines and that due to the continuum. As the MRS data is strongly affected by nuclear point source effects, and mitigation still leaves strong artifacts, we introduced an improved method which can be used when both imaging and MRS data are available which allows recovery of the contamination with strong mitigation of PSF effects, while taking advantage of the better resolution offered by the imager over the IFU. We use this as a baseline to evaluate an estimation method which uses VLT/SINFONI and Spitzer/IRS spectroscopy, to model line contamination in objects which only have imaging data (building upon the method of Haidar et al. 2024, as described in Section 3.4). This allows us to estimate the contamination present, and draw qualitative conclusions about the source of the extended MIR emission without the requirement of expensive MRS data. We have explored the caveats of each method. We note small-scale variations in contamination levels between the methods (of the order $\sim 5 - 10\%$), and as such, we caution against strong physical interpretation of variations on these scales. We demonstrate that the methods can be relied upon for conclusions regarding the presence and morphology of extended line-free (mostly continuum) emission.

Through the application of our methods and the further analysis, we find:

- Contamination from emission lines is important and should not be discounted in the study of central structure in broadband images of nearby AGN and galaxies.
- Contamination levels vary drastically across our sample, and so cannot be estimated across a population and must be treated explicitly for each object.
- Where possible, using both imaging and MRS together offers the best method to measure the contamination, however having both datasets available is expensive in terms of observing time.
- If only MRS or imaging data is available, the contamination may still be quantified – we recommend imaging data combined with ancillary data such as SINFONI in our estimation method, as this slightly overestimates the contamination (by approximately 5–10%), compared to using only MRS data which can both overestimate or underestimate the contamination by the same margin.
- After removal of emission line contamination, we find widespread evidence for extended MIR structures on $\sim 100$ pc scales, regardless of the method used.
- Among MIRI filters in local galaxies, contamination levels are highest in the F1000W band, and become less pronounced in longer wavelengths. However the longer wavelength bands suffer from poorer resolution, and so the recovery of the structures in F1000W offers the best trade-off between sensitivity and resolution.

This paper exists as a companion and continuation of Rosario et al. (prep), which introduced the sample and identified the MIR extensions. We present the best estimates of contamination in the objects in the sample across the filters used for readers to easily identify extended continuum emission. This work is built upon further in Lopez-Rodriguez et al. (prep), where the physical properties of the extended continuum emission are explored, and in Haidar et al. (prep), where the line-free MIR morphologies are compared to other phases, such as those traced by radio synchrotron emission. This paper also serves as a guide to best practice for addressing the issues of emission line contamination in the mid-infrared. Whether this method can also be applied to other JWST datasets (eg. NIRCam and NIRSpec), or other datasets more widely, is a fruitful avenue worth further exploration.

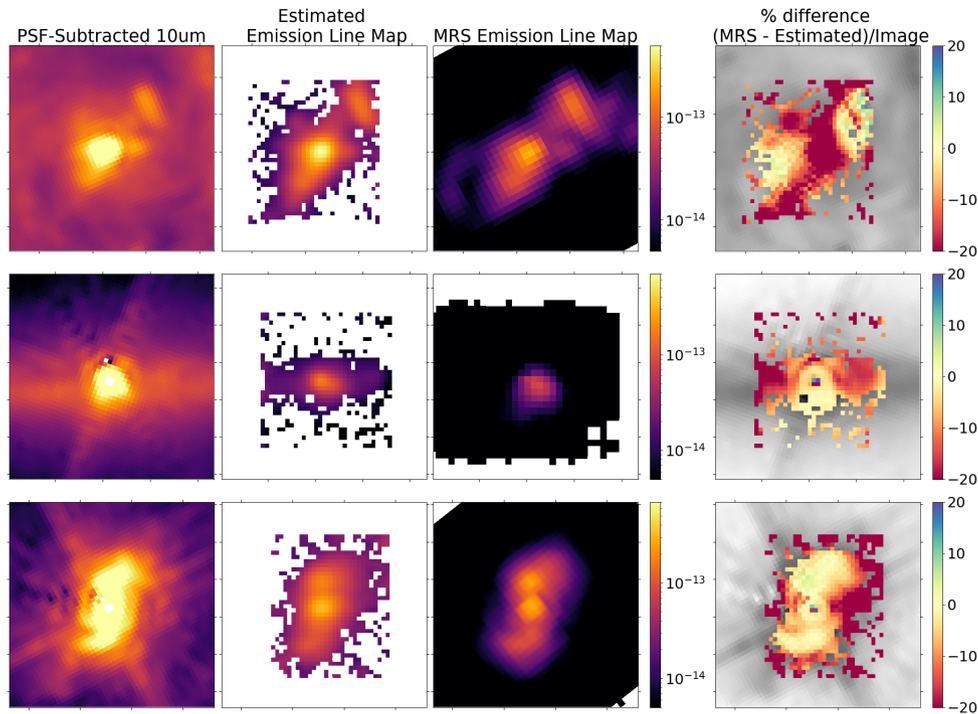

**Figure 6.** Comparison of the linemaps for the three shared objects (top: NGC 5728, middle: NGC 7172, bottom: NGC 3081) generated from mock maps from SINFONI (second column) to those measured straight from MRS (third column). The fourth column shows the percentage difference between these methods.

## ACKNOWLEDGEMENTS

DJR, HH, and SC acknowledge the support of the UK's Science and Technology Facilities Council (STFC) through grant ST/X001105/1. CRA and AA acknowledge support from the Agencia Estatal de Investigación of the Ministerio de Ciencia, Innovación y Universidades (MCIU/AEI) under the grant "Tracking active galactic nuclei feedback from parsec to kiloparsec scales", with reference PID2022−141105NB−I00 and the European Regional Development Fund (ERDF). AA also acknowledges financial support by the European Union grant WIDERA ExGal-Twin, GA 101158446. AAH and LHM acknowledge support from grant PID2021-124665NB-I00 funded by MCIN/AEI/10.13039/501100011033 and by ERDF A way of making Europe. SFH acknowledges support through UKRI/STFC grants ST/Y001656/1 and ST/V001000/1. E.B. acknowledges support from the Spanish grants PID2022-138621NB-I00 and PID2021-123417OB-I00, funded by MCIN/AEI/10.13039/501100011033/FEDER, EU. This work is based [in part] on observations made with the NASA/ESA/CSA James Webb Space Telescope. This research has made use of the NASA/IPAC Extragalactic Database (NED), which is operated by the Jet Propulsion Laboratory, California Institute of Technology, under contract with the National Aeronautics and Space Administration. This work makes use of several PYTHON packages: NUMPY (Harris et al. 2020), MATPLOTLIB (Hunter 2007), and ASTROPY (Astropy Collaboration et al. 2013, 2022).

## DATA AVAILABILITY

All observations in this paper are publicly available in archives with the proposal IDs given in Section 2, and can be processed as described in this paper. Specific analysis products from the paper are available on request.

## REFERENCES

Abuter R., Schreiber J., Eisenhauer F., Ott T., Horrobin M., Gillesen S., 2006, New Astron. Rev., 50, 398
Alonso-Herrero A., et al., 2011, ApJ, 736, 82
Alonso-Herrero A., et al., 2018, ApJ, 859, 144
Alonso-Herrero A., et al., 2019, A&A, 628, A65
Alonso-Herrero A., et al., 2021, A&A, 652, A99
Alonso Herrero A., et al., 2023, A&A, 675, A88
Antonucci R., 1993, ARA&A, 31, 473
Argyriou I., et al., 2023, A&A, 675, A111
Asmus D., 2019, MNRAS, 489, 2177
Asmus D., Hönig S. F., Gandhi P., 2016, ApJ, 822, 109
Astropy Collaboration et al., 2013, A&A, 558, A33
Astropy Collaboration et al., 2018, AJ, 156, 123
Astropy Collaboration et al., 2022, ApJ, 935, 167
Audibert A., Riffel R., Sales D. A., Pastoriza M. G., Ruschel-Dutra D., 2017, MNRAS, 464, 2139
Baumgartner W. H., Tueller J., Markwardt C. B., Skinner G. K., Barthelmy S., Mushotzky R. F., Evans P. A., Gehrels N., 2013, ApJS, 207, 19
Black J. H., van Dishoeck E. F., 1987, ApJ, 322, 412

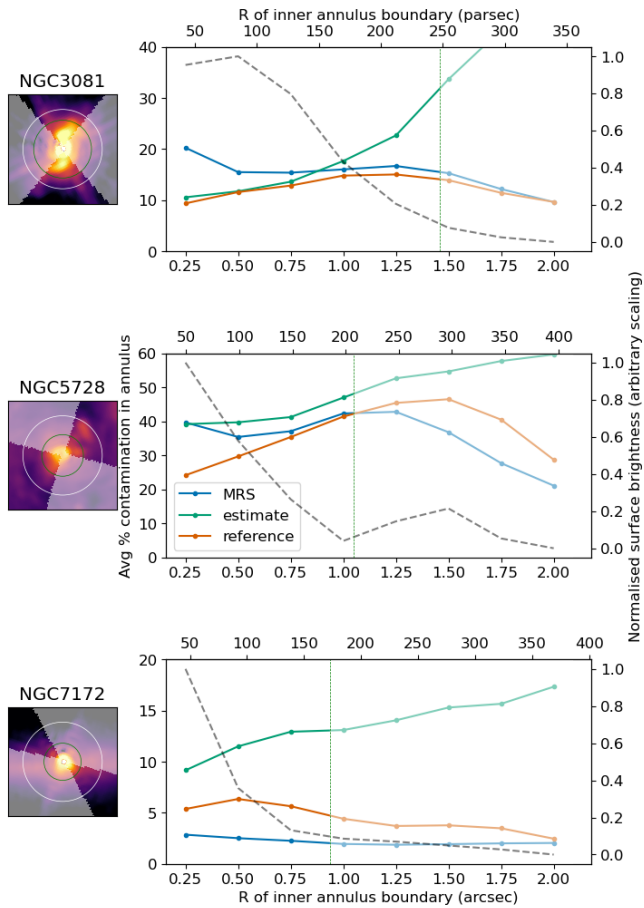

**Figure 7.** Radial average contamination percentage measured within sector covered by ionisation cone. On the left, we show thumbnails of the MIRI 10μm images, with the regions outside the ionisation cone masked. The white circle indicates the 2 arcsec radius covered by the radial plot, and the green circle shows the radius at which the average flux reaches 10% of the peak

Bock J. J., Marsh K. A., Ressler M. E., Werner M. W., 1998, ApJ, 504, L5
Bonnet H., et al., 2004, The Messenger, 117, 17
Burtscher L., et al., 2016, A&A, 586, A28
Cameron M., Storey J. W. V., Rotaciuc V., Genzel R., Verstraete L., Drapatz S., Siebenmorgen R., Lee T. J., 1993, ApJ, 419, 136
Combes F., et al., 2019, A&A, 623, A79
D'Agostino J. J., et al., 2019, MNRAS, 487, 4153
Davies R., et al., 2011, ApJ, 741, 69
Davies R., et al., 2024, A&A, 689, A263
Delaney D., Berger C., Hicks E., Burtscher L., Rosario D., Müller-Sánchez F., Malkan M., 2025, ApJ, 984, 163
Di Matteo T., Springel V., Hernquist L., 2005, Nature, 433, 604
Dicken D., et al., 2024, A&A, 689, A5
Eisenhauer F., et al., 2003, in Iye M., Moorwood A. F. M., eds, Society of Photo-Optical Instrumentation Engineers (SPIE) Conference Series Vol. 4841, Instrument Design and Performance for Optical/Infrared Ground-based Telescopes. pp 1548–1561 (arXiv:astro-ph/0306191), doi:10.1117/12.459468
Elitzur M., 2012, ApJ, 747, L33
Esparza-Arredondo D., et al., 2019, ApJ, 886, 125
Esparza-Arredondo D., Gonzalez-Martín O., Dultzin D., Masegosa J., Ramos-Almeida C., García-Bernete I., Fritz J., Osorio-Clavijo N., 2021, A&A, 651, A91
Esparza-Arredondo D., et al., 2025, A&A, 693, A174
Fabian A. C., 2012, ARA&A, 50, 455
GRAVITY Collaboration et al., 2020a, A&A, 634, A1
GRAVITY Collaboration et al., 2020b, A&A, 635, A92
García-Bernete I., et al., 2016, MNRAS, 463, 3531
García-Bernete I., et al., 2019, MNRAS, 486, 4917
García-Bernete I., et al., 2021, A&A, 645, A21
García-Bernete I., et al., 2022a, A&A, 666, L5
García-Bernete I., et al., 2022b, A&A, 667, A140
García-Bernete I., et al., 2024a, arXiv e-prints, p. arXiv:2409.05686
García-Bernete I., et al., 2024b, A&A, 681, L7
García-Burillo S., et al., 2016, ApJ, 823, L12
García-Burillo S., et al., 2019, A&A, 632, A61
García-Burillo S., et al., 2021, A&A, 652, A98
Gardner J. P., et al., 2023, PASP, 135, 068001
Gelbord J. M., Mullaney J. R., Ward M. J., 2009, MNRAS, 397, 172
González-Martín O., et al., 2019a, ApJ, 884, 10
González-Martín O., et al., 2019b, ApJ, 884, 11
González-Martín O., et al., 2023, A&A, 676, A73
González-Martín O., et al., 2025, MNRAS, 539, 2158
Gravity Collaboration et al., 2024, A&A, 690, A76
Haidar H., et al., 2024, arXiv e-prints, p. arXiv:2404.16100
Haidar H., et al., in prep
Harris C. R., et al., 2020, Nature, 585, 357
Harrison C. M., Ramos Almeida C., 2024, Galaxies, 12, 17
Heckman T. M., Best P. N., 2014, ARA&A, 52, 589
Helou G., Madore B. F., Schmitz M., Bicay M. D., Wu X., Bennett J., 1991, in Albrecht M. A., Egret D., eds, Astrophysics and Space Science Library Vol. 171, Databases and On-line Data in Astronomy. pp 89–106, doi:10.1007/978-94-011-3250-3_10
Hermosa Muñoz L., et al., 2024, A&A, 690, A350
Hönig S. F., 2019, ApJ, 884, 171
Hönig S. F., Kishimoto M., 2017, ApJ, 838, L20
Hönig S. F., Kishimoto M., Antonucci R., Marconi A., Prieto M. A., Tristram K., Weigelt G., 2012, ApJ, 755, 149
Houck J. R., et al., 2004, ApJS, 154, 18
Hunter J. D., 2007, Computing in Science and Engineering, 9, 90
Labiano A., et al., 2016, in Peck A. B., Seaman R. L., Benn C. R., eds, Society of Photo-Optical Instrumentation Engineers (SPIE) Conference Series Vol. 9910, Observatory Operations: Strategies, Processes, and Systems VI. p. 99102W (arXiv:1608.05312), doi:10.1117/12.2232554
Lebouteiller V., Barry D. J., Spoon H. W. W., Bernard-Salas J., Sloan G. C., Houck J. R., Weedman D. W., 2011, ApJS, 196, 8
Leist M. T., et al., 2024, AJ, 167, 96
Lopez-Rodriguez E., et al., in prep
Mattson B. J., Weaver K. A., 2004, ApJ, 601, 771
May D., Rodríguez-Ardila A., Prieto M. A., Fernández-Ontiveros J. A., Diaz Y., Mazzalay X., 2018, MNRAS, 481, L105
McKinney J., et al., 2023, ApJ, 946, L39
Morganti R., 2017, Frontiers in Astronomy and Space Sciences, 4, 42
Osterbrock D. E., 1981, ApJ, 249, 462
Pereira-Santaella M., et al., 2022, A&A, 665, L11
Perrin M. D., Soummer R., Elliott E. M., Lallo M. D., Sivaramakrishnan A., 2012, in Space Telescopes and Instrumentation 2012: Optical, Infrared, and Millimeter Wave. SPIE, p. 84423D, doi:10.1117/12.925230
Prieto M. A., Mezcua M., Fernández-Ontiveros J. A., Schartmann M., 2014, MNRAS, 442, 2145
Radomski J. T., Piña R. K., Packham C., Telesco C. M., De Buizer J. M., Fisher R. S., Robinson A., 2003, ApJ, 587, 117
Ramos Almeida C., Ricci C., 2017, Nature Astronomy, 1, 679
Ramos Almeida C., et al., 2009, ApJ, 702, 1127
Ramos Almeida C., et al., 2011, ApJ, 731, 92
Reyes-Amador O. U., et al., 2024, MNRAS, 531, 1841
Robitaille T., et al., 2024, astropy/reproject: v0.13.1, doi:10.5281/zenodo.10931886, https://doi.org/10.5281/zenodo.10931886
Rosario D., et al., in prep
Sarangi A., Dwek E., Kazanas D., 2019, ApJ, 885, 126
Silk J., Rees M. J., 1998, A&A, 331, L1
Soliman N. H., Hopkins P. F., 2023, MNRAS, 525, 2668

Sturm E., Lutz D., Verma A., Netzer H., Sternberg A., Moorwood A. F. M., Oliva E., Genzel R., 2002, A&A, 393, 821
Tristram K. R. W., Burtscher L., Jaffe W., Meisenheimer K., Hönig S. F., Kishimoto M., Schartmann M., Weigelt G., 2014, A&A, 563, A82
Urry C. M., Padovani P., 1995, PASP, 107, 803
Veilleux S., Cecil G., Bland-Hawthorn J., 2005, ARA&A, 43, 769
Veilleux S., Maiolino R., Bolatto A. D., Aalto S., 2020, A&ARv, 28, 2
Ward S. R., Costa T., Harrison C. M., Mainieri V., 2024, MNRAS, 533, 1733
Wells M., et al., 2015, PASP, 127, 646
Werner M. W., et al., 2004, ApJS, 154, 1
Williamson D., Hönig S., Venanzi M., 2020, ApJ, 897, 26
Wright G. S., et al., 2023, PASP, 135, 048003
Zhang L., et al., 2024a, ApJ, 974, 195
Zhang L., et al., 2024b, ApJ, 975, L2

## APPENDIX A: COMPARISON OF MRS LINEMAPS TO TRACER LINEMAPS FROM SINFONI

The linemaps of the proxy species used from SINFONI are shown alongside linemaps from the MRS data of the true contaminants in F1000W for NGC 5728 in Figure A1. The coloured boxes indicate which species are used as proxies for one another. While there are differences, the overall morphology is qualitatively traced well.

## APPENDIX B: RESULTS WITHOUT MRS PSF MITIGATION

In Section 3.2.1, we describe the necessary step of mitigating the bright point source PSF in the MRS data. To illustrate the effects of this, we include below the results of our method without this process to be compared with the MRS results presented in Figure 4.

## APPENDIX C: QUANTIFICATION OF LINE CONTAMINATION FOR THE SAMPLES

This appendix presents the contamination maps alongside the line-free images for all the objects in the samples which are not shared between the MRS sample and imaging sample. This is to allow quick qualitative judgments about contamination to be made by readers who are interested in potential extended MIR emission in these objects which may have been seen in the initial MIRI images in Rosario et al. (prep), in order to facilitate future science. The objects here are only represented by either imaging or MRS, and thus the optimal combined method could not be applied – the methods used for these are described in the "MRS only" (Section 3.2) or "imaging-only" (Section 3.4) sections.

### C1 Objects with only MRS Data

For these objects, the MRS-only method (described in Section 3.2) is applied across all filters.

### C2 Objects with only Imaging Data

For these objects, the "imaging-only" estimation method is applied, and varies slightly across each filter to account for the different contaminating species. All filters above F1000W are fully covered by Spitzer/IRS data, so the scaling for the mock maps can be done by direct flux measurements as with the [S IV] line in F1000W. Below we outline the choices of tracers used.

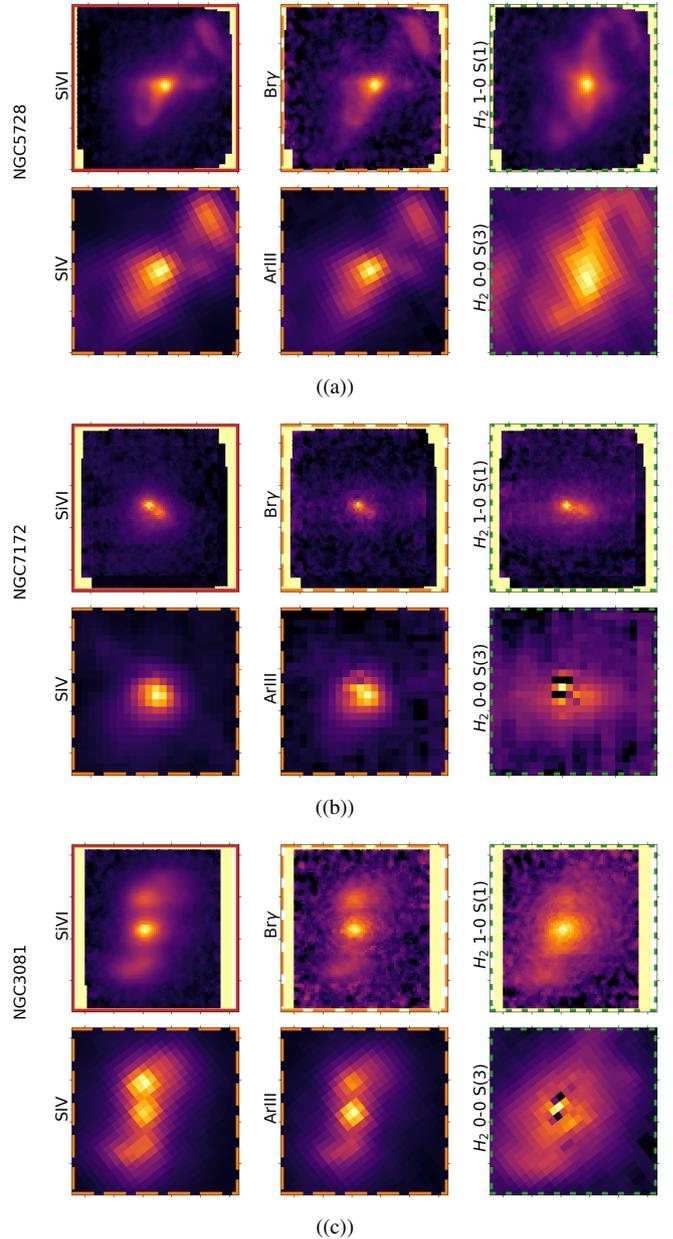

**Figure A1.** Linemaps for each of the shared objects, showing the lines used from SINFONI (top row) alongside the lines which are contaminating F1000W (bottom row) from MRS data. The tracer which is used for each contaminating line are shown by the colored dashed lines framing the linemaps. These maps are normalised by their total flux, and shown with a square-root stretch so as to allow visual comparison of the morphologies in order to assess the suitability of each species as a morphological tracer for the other. We note the artifacts present in the $H_2$ 0-0 S(3) maps derived from the MRS data, which are due to the weak line being overshadowed by the "wiggles" in the continuum caused by instrument effects. These maps are not used directly at any point of this work, and are shown here for qualitative morphological comparison only.

**F1500W:** This filter is contaminated by [Ne v] and [Ne III]. With an ionisation potential of 97.1eV, we elect to trace [Ne v] using [Si VI] from SINFONI; and with an ionisation potential of 41eV, we elect to trace [Ne III] using Brγ.

**F1800W:** This filter is contaminated by $H_2$ 0-0 S(1) and [S III]. With an ionisation potential of 23.3eV, we elect to trace [S III] using

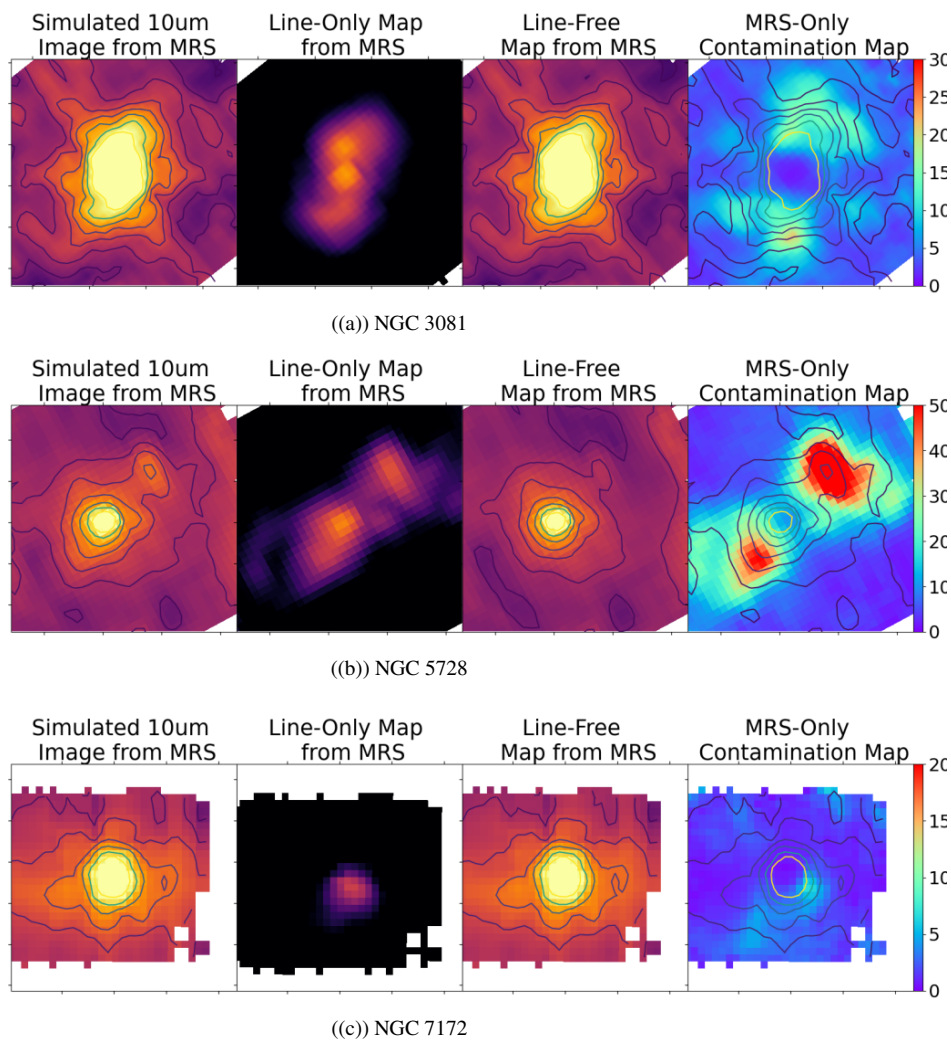

**Figure B1.** Results of using the MRS-only method without mitigation of the PSF on the three shared objects in F1000W – NGC 3081, NGC 5728, NGC 7172. From left to right, the full image is synthesised from the MRS data, and can be decomposed into line-only emission and line-free emission. In this case, the synthetic line-free map is effectively the decontaminated image. The contamination map is the percentage of flux in the line-free map compared with the synthetic image. The contours in these images, in all cases, represent the flux. The contours in the first column shows the flux in the image. The contours in the third column are that of the flux in the decontaminated map. The contours in the fourth column are those of the flux in the initial image.

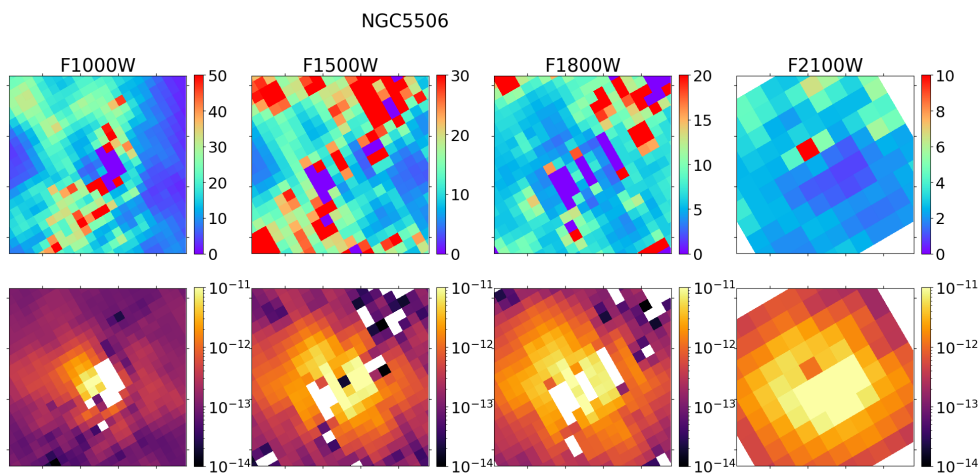

**Figure C1.** Contamination maps (top row) and line-free synthetic images (bottom row) shown across the 4 filters for NGC 5506. These were generated using the MRS-only method described in Section 3.2.

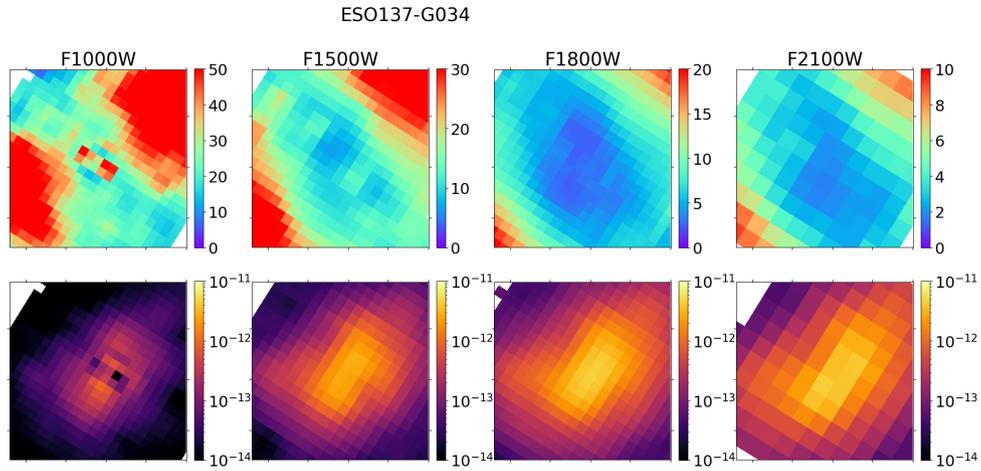

**Figure C2.** Contamination maps (top row) and line-free synthetic images (bottom row) shown across the 4 filters for ESO 137-G034. These were generated using the MRS-only method described in Section 3.2.

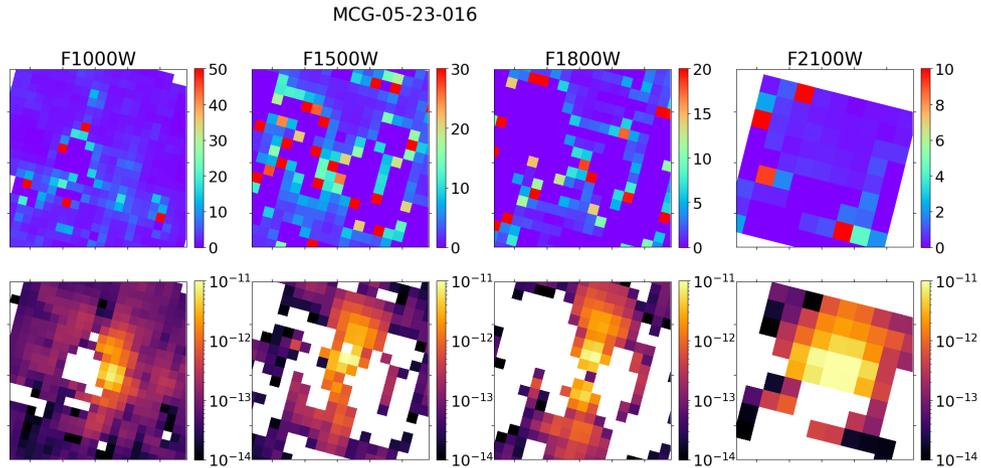

**Figure C3.** Contamination maps (top row) and line-free synthetic images (bottom row) shown across the 4 filters for MCG 05-23-016. These were generated using the MRS-only method described in Section 3.2.

Br$\gamma$ from SINFONI; and the molecular Hydrogen is traced by the molecular Hydrogen line found in SINFONI.

**F2100W:** This filter is contaminated by only [S III]. We trace this in the same way as in F1800W.

This paper has been typeset from a T$_E$X/L$^A$T$_E$X file prepared by the author.

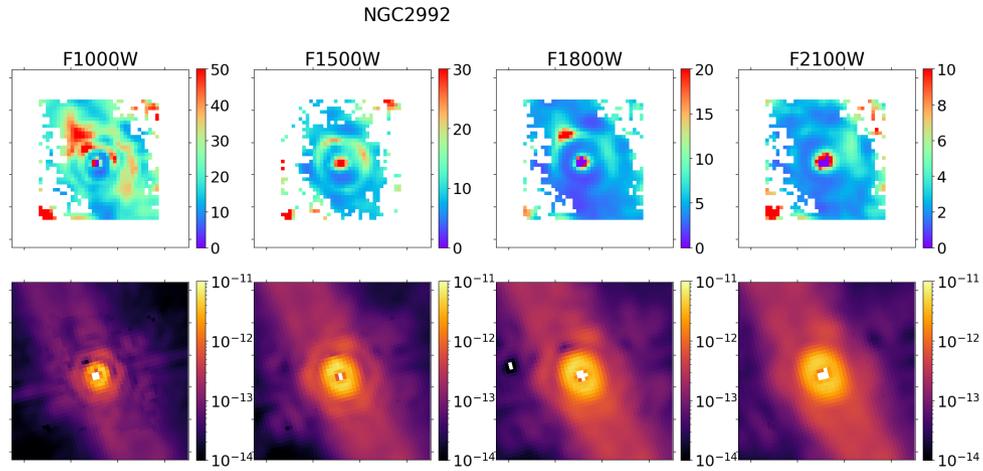

**Figure C4.** Contamination maps (top row) and line-free synthetic images (bottom row) shown across the 4 filters for NGC 2992. These were generated using the imaging-only estimation method described in Section 3.4, with the tracers used for the relevant lines in other bands described in Section C2.

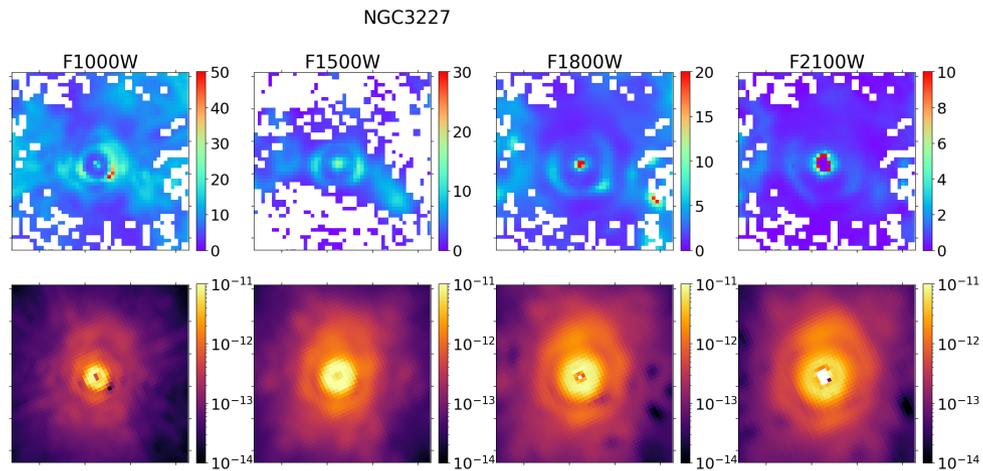

**Figure C5.** Contamination maps (top row) and line-free synthetic images (bottom row) shown across the 4 filters for NGC 3227. These were generated using the imaging-only estimation method described in Section 3.4, with the tracers used for the relevant lines in other bands described in Section C2.

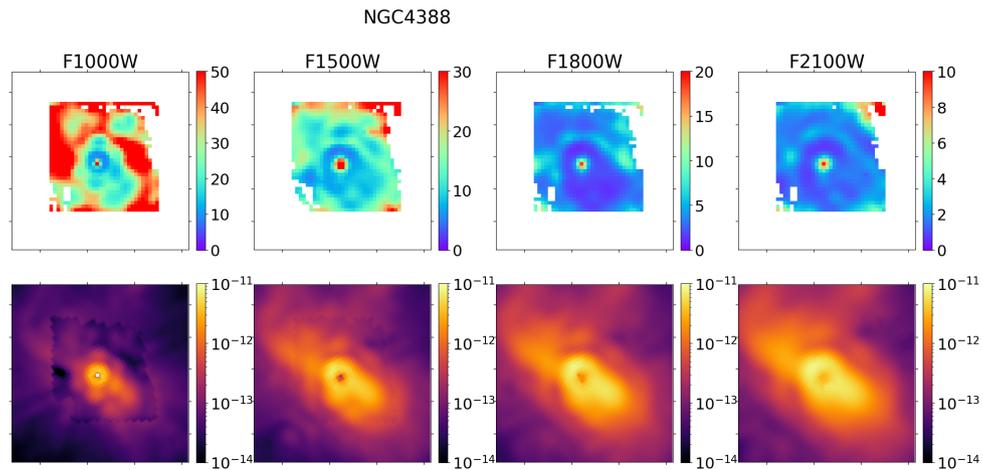

**Figure C6.** Contamination maps (top row) and line-free synthetic images (bottom row) shown across the 4 filters for NGC 4388. These were generated using the imaging-only estimation method described in Section 3.4, with the tracers used for the relevant lines in other bands described in Section C2.